\documentclass[a4paper,11pt]{article}
\pdfoutput=1 
\usepackage{jheppub,amsthm} 
\usepackage[utf8]{inputenc}
\usepackage{booktabs,multirow, quotes}
\usepackage{dsfont}
\usepackage[english]{babel}
\usepackage{amsmath,amssymb,graphicx,xcolor,alltt}
\usepackage{framed}

\newcommand{\beq}{\begin{equation}}
\newcommand{\eeq}{\end{equation}}
\newcommand{\bea}{\begin{eqnarray}}
\newcommand{\ea}{\end{eqnarray}}

\def\be{\begin{equation}}
\def\ee{\end{equation}}

\makeatletter
\def\@fpheader{\ }
\makeatother

\title{\boldmath $\mathcal{N}=2$ RG flows, Non-Invertible Symmetries and Matrix Factorisations}

\author[a]{Federico Ambrosino,}
\author[b]{Matthias R.\ Gaberdiel,}
\author[c]{and Yu Nakayama}

\affiliation[a]{Perimeter Institute for Theoretical Physics, Waterloo, Ontario N2L 2Y5, Canada}
\affiliation[b]{Institut f\"ur Theoretische Physik, ETH Zurich, CH-8093 Z\"urich, Switzerland}
\affiliation[c]{Yukawa Institute for Theoretical Physics, Kyoto University, Kitashirakawa Oiwakecho, Sakyo-ku, Kyoto 606-8502 Japan}

\emailAdd{federicoambrosino25@gmail.com}
\emailAdd{gaberdiel@itp.phys.ethz.ch}
\emailAdd{yu.nakayama@yukawa.kyoto-u.ac.jp}

\hypersetup{
  pdftitle  = {N=2 RG flows, Non-Invertible Symmetries and Matrix Factorisations},
  pdfauthor = {Federico Ambrosino, Matthias R. Gaberdiel, and Yu Nakayama},
  pdfsubject = {Topological defects and RG flows in N=2 minimal models},
  pdfkeywords = {non-invertible symmetry, topological defects, RG flows,
                 matrix factorisations, N=2 minimal models}
}

\abstract{We study the behaviour of the topological defect lines of the $k^{\rm th}$ ${\cal N}=2$ minimal models that are preserved by the least relevant perturbation to first order.  It is usually believed that these defects should then also define symmetries of the IR theory, which for the usual ``massless'' flow should be the $(k-2)^{\rm nd}$ ${\cal N}=2$ minimal model. Using CFT arguments we show that this is not possible.  We also reproduce this result using matrix factorisation techniques: while the corresponding B-type defects can be adjusted to first order in the deformation, there is an obstruction at second order, which is associated with a supersymmetry anomaly. By contrast, for the associated massive integrable flow, which corresponds to a Chebyshev deformation of the superpotential, all of these defects can be consistently deformed, and they indeed define symmetries of the massive IR theory.}

\begin{document}

\begin{flushright}
YITP-26-94
\\
\end{flushright}
\maketitle

\newpage

\section{Introduction and Summary}
Symmetry is one of our main tools for understanding a renormalisation group (RG) flow without solving it. A symmetry of the ultraviolet theory that is preserved along an RG flow must also be a symmetry of the infrared fixed point, and this constrains the possible RG flows severely. 
In recent years the notion of symmetry entering this statement has been broadened considerably, see e.g.\ \cite{Schafer-Nameki:2023jdn,Shao:2023gho} for recent reviews: a symmetry can now be a topological defect, an extended operator whose correlators are invariant under deformations of its shape, and it need not be invertible. Instead of a group law, such defects obey fusion rules~\cite{Oshikawa:1996dj,Fuchs:2002cm, Frohlich:2006ch}. 
In two dimensions the relevant symmetries are often generated by topological defect lines, and they are expected to remain symmetries provided that the local operator by which the action is deformed commutes with the line.  
If this is the case, the corresponding non-invertible line must also act on the infrared fixed point 
\cite{Fredenhagen:2009tn,Gaiotto:2012np,Cordova:2024vsq,Komargodski:2020mxz,Chang:2018iay}, and this reasoning underlies a recent set of conjectured RG flows~\cite{Nakayama:2024msv,Ambrosino:2025yug,Gaberdiel:2026sfg, Ambrosino:2026umb, Benedetti:2026drn}. In this paper we show that this expectation can fail, and we trace the mechanism in an exactly solvable $\mathcal{N}=2$ model.

Our example is the two-dimensional $\mathcal{N}=2$ Landau--Ginzburg model with superpotential $W=X^{k+2}$, which describes the $k^{\rm th}$ $\mathcal{N}=2$ minimal model \cite{Martinec:1988zu,Vafa:1988uu}. Among its topological defect lines is a family of non-invertible ones, the integer-spin Verlinde lines, which form the fusion category $\mathfrak{so}(3)_k$. We deform the superpotential by its least relevant chiral perturbation, $W=X^{k+2}+\lambda X^k$. Since $\mathcal{N}=2$ supersymmetry protects the superpotential from corrections, the massless sector of the infrared is the $\mathcal{N}=2$ minimal model with $W=X^k$, that is, the $(k-2)^{\rm nd}$ minimal model. In addition, the deformed superpotential has two non-degenerate critical points at $X\sim\pm i\sqrt{\lambda}$, and these describe two massive vacua to which we return below. The perturbing field commutes with every line of $\mathfrak{so}(3)_k$, so one would expect that all of $\mathfrak{so}(3)_k$ survives as a symmetry of the infrared theory. This, however, cannot hold in its simplest form. The $(k-2)^{\rm nd}$ minimal model is a smaller theory, and its fusion category contains no subcategory isomorphic to $\mathfrak{so}(3)_k$ (we make this statement precise in Section~\ref{sec:puzzle}). This is the puzzle we analyse.

The issue is not a minor one, since the assumption that a deformation commuting with a symmetry preserves it in the infrared is central to the symmetry-based approach to RG flows~\cite{Gaiotto:2012np,Chang:2018iay,Nakayama:2024msv,Ambrosino:2025yug,Gaberdiel:2026sfg, Ambrosino:2026umb, Benedetti:2026drn}. Our analysis exposes two distinct subtleties. 
The first is that carrying a line through the flow is not automatic: beyond the leading order the line has to be dressed by defect-localised counterterms, and this dressing may fail, since there may be an anomaly localised on the line. 
The second subtlety is that the infrared theory may contain massive vacua in addition to the massless sector, so that it is not sufficient to analyse the symmetries on the massless part alone. The advantage of the $\mathcal{N}=2$ setting is that both phenomena can be exhibited exactly. The B-type defects, which are the topological defect lines compatible with the supersymmetric structure, are described by matrix factorisations of the superpotential~\cite{Kapustin:2002bi,Brunner:2003dc, Lazaroiu:2003zi}, and their fate under the deformation can be analysed systematically, order-by-order in $\lambda$. For the usual massless flow, we show that there is indeed an obstruction at second order in $\lambda$, and it gives rise to an anomaly with respect to supersymmetry. On the other hand, one can also modify the deformation so that the defects can be deformed consistently to all orders in $\lambda$. The associated RG flow is then an integrable flow whose end-point is a purely massive theory (and, in particular, not the $(k-2)^{\rm nd}$ minimal model).
\smallskip

We now describe the above topological defect lines in more detail, since they are the main objects of our analysis. Like the primaries, the defects of the $k^{\rm th}$ minimal model are labelled by triples $[L,M,S]$, and each acts on a given state by an eigenvalue determined by the modular $S$-matrix. The perturbing field lies in the sector $(0,-2,0)$. Since a Cardy line acts on each sector by a single eigenvalue, it commutes with the perturbation if and only if it acts as the identity on the perturbing field. This selects the lines with $M=0$. Restricting to the defects that commute with the full superconformal algebra ($S=0$) then forces $L$ to be even, and we are left with the family $[L,0,0]$, $L=0,2,4,\dots$. These close under fusion into the integer-spin subring of the $\mathfrak{su}(2)_k$ fusion ring, the non-invertible category $\mathfrak{so}(3)_k\cong \mathfrak{su}(2)_k^{\mathrm{even}}$. Most of these lines are genuinely non-invertible, with quantum dimensions
\be\label{qdim}
q[L,k] = \frac{\sin\!\big(\frac{\pi(L+1)}{k+2}\big)}{\sin\!\big(\frac{\pi}{k+2}\big)} > 1 \qquad (0<L<k) \ .
\ee
The fate of this family of lines along the flow is the subject of the paper, and the algebraic form of the puzzle is that 
this subcategory has no counterpart in the $(k-2)^{\rm nd}$ minimal model, see Section~\ref{sec:puzzle}.

In the Landau--Ginzburg description the defect lines are described by matrix factorisations of the superpotential $W=X^d-Y^d$, where $d=k+2$, and the two terms are copies of the bulk superpotential in the two variables $X$ and $Y$. A matrix factorisation is a pair of polynomial matrices $E,J$ (in the variables $X$ and $Y$), satisfying $EJ=JE=W\cdot {\bf 1}$. The simplest ones are obtained by writing $W$ as a product of linear factors and distributing them between $E$ and $J$, and the line $[2\ell,0,0]$ corresponds to one such choice. Preserving the line as a supersymmetric (B-type) defect along the flow amounts to deforming $E$ and $J$ order-by-order in $\lambda$ so that their product remains equal to the deformed superpotential. In this way the question of whether the symmetry is preserved, compatibly with supersymmetry, becomes the question of whether the matrix factorisation can be deformed. This can be answered by an explicit computation in a suitable polynomial ring. The behaviour of these topological defects under bulk perturbations has been studied before by means of RG defects \cite{Brunner:2020lrm}, with the conclusion that for generic flows only a few invertible defects and one bound state thereof reach the infrared. Our order-by-order analysis confirms this conclusion and explains in more detail where the obstruction comes from: it appears at second order, as an anomaly localised on the line.

This computation therefore reproduces the puzzle within the Landau--Ginzburg description: at first order in $\lambda$ the matrix factorisation can be deformed, as expected, since the line commutes with the perturbation. At second order it cannot, since there is an obstruction that no choice of first-order data removes. One might ask whether this is an artefact of insisting on the simplest, rank-1 description, and whether embedding the line in a larger matrix factorisation together with trivial defects would help. We show that it does not, and that the obstruction persists at the same order. (The same obstruction appears in the folded picture, where the line becomes a boundary condition and the deformed supercharge must square to the boundary superpotential. The anomaly is precisely the impossibility of correcting the supercharge beyond first order, see Section~\ref{sec:BRST}.) No supersymmetry-compatible realisation of the non-invertible lines therefore survives beyond first order along the deformation $W=X^{k+2}+\lambda X^k$. This is the matrix-factorisation counterpart of the fusion-ring obstruction we started from. Whether the lines survive in some form that is not compatible with the supersymmetric structure is a question that the matrix factorisation method leaves open.

There is nonetheless a way to preserve the symmetry if we allow the superpotential itself to be corrected in $\lambda$ beyond the linear term $\lambda X^k$ as a different choice of bare deformation. Within the natural ansatz in which each quadratic factor of the factorised superpotential difference is shifted by a constant, cancelling the obstruction order-by-order fixes the corrected superpotential uniquely. The result is the Chebyshev (Dickson) deformation, which takes the compact form $W(X,\lambda)=\xi^d+a^d/\xi^d$ in the uniformising variable $X=\xi+a/\xi$ with $a=-\lambda/d$. This is a known integrable deformation of the $\mathcal{N}=2$ minimal models~\cite{Fendley:1991ve}, but the flow it generates is massive rather than massless:  instead of flowing to the $(k-2)^{\rm nd}$ minimal model, it ends in $k+1$ massive vacua. Along this massive flow all the lines of $\mathfrak{so}(3)_k$ are preserved to all orders, together with their fusion ring. This then reproduces the integer-spin part of the $\mathfrak{su}(2)_k$ fusion ring, in agreement with Gepner's identification of the chiral ring of the Chebyshev superpotential with the $\mathfrak{su}(2)_k$ fusion ring~\cite{Gepner:1990gr}. The symmetry-preserving deformation therefore does exist, but the end-point of the flow is different from what we had assumed. 

If we insist, on the other hand, on the linear deformation $W=X^{k+2}+\lambda X^k$, then it is not possible to preserve the defect lines as supersymmetric (B-type) defects along the entire flow. There is, however, a natural way in which the defect lines \emph{can be deformed} if we relax the requirement that the associated matrix factorisation remains polynomial in the variables $X$ and $Y$, see the discussion in Section~\ref{sec:5.1}. If we follow the roots of the deformed superpotential into the infrared, the resulting defect also involves the two massive vacua at $X\sim\pm i\sqrt{\lambda}$ (in addition to the $(k-2)^{\rm nd}$ minimal model). This continuation is formal, in that the matrix factorisation is not polynomial in $X$ and $Y$ at intermediate values of $\lambda$, and we do not have a precise interpretation of it. It does, however, suggest that the obstruction may also be understood in terms of the additional massive vacua that accompany the $(k-2)^{\rm nd}$ minimal model. We discuss these questions, as well as related issues such as the behaviour of the closely related $\mathbb{Z}_k$ parafermion flow and integrability, in Section~\ref{sec:discussion}.
\smallskip

\paragraph{Outline.}
The paper is organised as follows. Section~\ref{sec:puzzle} introduces the $\mathcal{N}=2$ minimal models and their defects, and explains the puzzle in detail. Section~\ref{sec:MF} develops the matrix-factorisation description and establishes the second-order obstruction. Section~\ref{sec:massive} shows that consistency to all orders forces the Chebyshev deformation, and discusses flat coordinates and integrability. Section~\ref{sec:discussion} is devoted to the physical interpretation of the obstruction. More technical proofs are given in Appendices~\ref{app:rank}--\ref{app:even}.

\section{The CFT Puzzle}\label{sec:puzzle}
In this paper we study the behaviour of the defects in $\mathcal{N}=2$ minimal models under suitable perturbations.

\subsection[\texorpdfstring{$\mathcal{N}=2$}{N=2} preliminaries]{\boldmath \texorpdfstring{$\mathcal{N}=2$}{N=2} preliminaries}\label{sec:prelim}

Let us begin by fixing our notation. The $\mathcal{N}=2$ minimal models have central charge 
\be
c=\frac{3k}{k+2}\ ,
\ee
where $k=1,2,\ldots$. They are most conveniently described by the coset construction, see e.g.\ \cite{DiFrancescoMathieuSenechal1997}
\be
\bigl(\mathcal{N}=2 \bigr)_k = \frac{\mathfrak{su}(2)_k \oplus \mathfrak{so}(2)_1}{\mathfrak{u}(1)} \ , 
\ee
where $\mathfrak{so}(2)_1$ describes two free fermions. The different irreducible representations are then labelled by three integers $(l,m,s)$, where
$l=0,\dotsc,k$, and $m$ and $s$ are defined modulo $2 (k+2)$ and $4$,
respectively. Here $l+m+s$ is even, and we have the field
identifications $(l,m,s)\sim(k-l,m+k+2,s+2)$. These sectors are
representations of the bosonic subalgebra of the $\mathcal{N}=2$ superconformal
algebra, and the conformal weight and $\mathfrak{u}(1)$ charge of the highest weight states are equal to 
\begin{align}
& h(l,m,s) = \frac{l(l+2) - m^2}{4(k+2)}  + \frac{s^2}{8} \ \ \hbox{mod $(1)$} \\
& q(l,m,s) = \frac{s}{2} - \frac{m}{k+2}  \ \ \hbox{mod $(1)$} \ . 
\end{align}
We shall work with the usual charge conjugation modular invariant, for which the full spectrum takes the form 
\be\label{fullspec}
\mathcal{H} = \bigoplus_{[l,m,s]} \mathcal{H}_{(l,m,s)} \otimes \overline{\mathcal{H}_{(l,-m,-s)}} \ , 
\ee
where the sum runs over all the representations modulo field identification. Then the $\mathcal{N}=2$ preserving defects are determined by the Cardy construction \cite{Cardy:1989ir}, see also \cite{Petkova:2000ip,Frohlich:2006ch}, and are hence labelled by the irreducible representations themselves. To distinguish them from the local fields, we shall use the notation $\mathcal{D}_{[L,M,S]}$ for these topological defects. 

The key ingredient in the Cardy construction is the $S$-matrix, which factorises as 
\be\label{Smatrix}
S_{(l,m,s),(l',m',s')} = {\frac{1}{\sqrt{2}(k+2)}}
\sin\left(\frac{\pi (l+1)(l'+1)}{k+2}\right)
\exp\left(\frac{i\pi m m'}{k+2}\right)
\exp\left(-\frac{i\pi s s'}{2}\right) 
\ee
into the $S$-matrix associated with $\mathfrak{su}(2)_k$ and the two $\mathfrak{u}(1)$ factors, respectively. In particular, the $S$-matrix determines the action of the defect $\mathcal{D}_{[L,M,S]}$ on any state in the sector associated with $[l,m,s]$ in (\ref{fullspec}):  it acts as multiplication by 
\be
\left. \gamma_{[L,M,S]} \right|_{[l,m,s]} = \frac{S_{(L,M,S),(l,m,s)}}{S_{(0,0,0),(l,m,s)}} \, {\bf 1}\ . 
\ee
Furthermore, the multiplication of defects is described by the fusion rules
\be
\mathcal{D}_{[L_1,M_1,S_1]} \cdot \mathcal{D}_{[L_2,M_2,S_2]} = 
\sum_{(L,M,S)} \mathcal{N}_{(L_1,M_1,S_1), (L_2,M_2,S_2)}{}^{(L,M,S)}\, \mathcal{D}_{[L,M,S]} \ , 
\ee
that are determined via the Verlinde formula \cite{Verlinde:1988sn} from the $S$-matrix. 

\subsection{The least relevant deformation}\label{sec:deform}

We are interested in the deformation that induces the flow from the $k^{\rm th}$ minimal model to the $(k-2)^{\rm nd}$. It is believed to be triggered by the superconformal descendant of the chiral primary $(k,k,0)$, which after field identification can also be written as 
\be
(k,k,2) \sim (0, -2, 0) \ . 
\ee
The defects $\mathcal{D}_{[L,M,S]}$ that are preserved by this perturbation are characterised by the condition that they commute with the action of the perturbing field, i.e.\ by the requirement 
\be
\mathcal{D}_{[L,M,S]} \, \Phi(z,\bar{z}) \, |\Psi\rangle = \Phi(z,\bar{z})\, \mathcal{D}_{[L,M,S]}  \,  |\Psi\rangle \ , 
\ee
where $\Phi(z,\bar{z})$ is the perturbing field, while $|\Psi\rangle$ is any state in $\mathcal{H}$. If the perturbing field is from the sector $(l,m,s)$, this is equivalent to the condition that 
\be
\gamma_{[L,M,S]}(l,m,s) =  \gamma_{[L,M,S]}(0,0,0)  \ .
\ee
Another way of saying this is that the defect acts on the perturbing field $(l,m,s)$ as it acts on the identity, i.e.\ by its quantum dimension. For the case at hand, $(l,m,s) = (0,-2,0)$, and thus it follows from 
eq.~(\ref{Smatrix}) that the defects with $M=0$ will have this property. In order to avoid the subtleties associated with the fermionic nature of the model, we shall furthermore assume that $S=0$, i.e.\ that the defect commutes with the full superconformal algebra (rather than just its bosonic subalgebra) \cite{Gaberdiel:2026sfg}. Thus the defects that commute with the above deformation include at least all the defects of the form $[L,0,0]$ where $L$ is even. 
Following the logic of \cite{Nakayama:2024msv,Ambrosino:2025yug,Gaberdiel:2026sfg,Ambrosino:2026umb,Benedetti:2026drn} one would therefore expect that an isomorphic ring of defects exists in the IR theory.

However, we now show that this is not possible. It follows from the structure of the $S$-matrix and the Verlinde formula that the defects of the form $[L,M=0,0]$ with $L$ even form a closed subring of the fusion ring, which is simply isomorphic to the integer spin subalgebra of the fusion ring of $\mathfrak{su}(2)_k$, i.e.\ to the ring 
\be\label{defectring}
[L_1] \otimes [L_2] = \bigoplus_{L = |L_1-L_2|}^{\min(L_1+L_2,2k - L_1 - L_2)} [L] \ , 
\ee
where the sum runs in steps of $2$. However, in the IR the value of $k$ is reduced by $2$, $k\mapsto k-2$, and it is relatively straightforward to see that, with a single exception, 
no such family of topological lines can exist in the IR theory. More explicitly, the image of the $[L=2]$ generator of the $k$-theory must be a linear combination of simple lines in the $(k-2)$-theory, 
\be
[L=2]_{k} \mapsto \bigoplus_j \, [L_j]_{k-2} \ , 
\ee
and the fact that the fusion rules are preserved by the flow to the IR means that the quantum dimensions of eq.~(\ref{qdim}) must agree, i.e.\ that 
\be\label{cond}
q[2,k] = \sum_{j} q[L_j,k-2] \ .
\ee
It follows directly from eq.~(\ref{qdim}) that $q[2,k]  \leq 3$, while $q[L_j,k-2]\geq 1$ for any allowed $0\leq L_j \leq k-2$. Thus, the sum over $j$ can run over at most $3$ terms, and for $k\geq 10$, the only possible terms that can appear are 
\begin{align}
q[0,k-2] & = q[k-2,k-2] = 1 \ , \\
q[1,k-2] & = q[k-3,k-2] = 2\cos\!\big(\tfrac{\pi}{k}\big) \leq 3 \ , \\
q[2,k-2] & = q[k-4,k-2] = 1+2\cos\!\big(\tfrac{2\pi}{k}\big) \leq 3 \ ,
\end{align}
since all other quantum dimensions are bigger than $3$. Thus there are only finitely many possibilities, and one finds that no combination of generators in the $(k-2)$ theory satisfies eq.~(\ref{cond}) for $k\geq 10$. A direct check also shows that there is no such solution for $k \leq 9$, with the single exception of $k=4$.\footnote{\label{fn:k4}At $k=4$, eq.~(\ref{cond}) admits the solution $q[2,4]=2=q[0,2]+q[2,2]$, and indeed $[0]\mapsto[0]$, $[2]\mapsto[0]\oplus[2]$, $[4]\mapsto[2]$ defines a fusion-ring homomorphism, namely the restriction ${\rm Rep}(S_3)\to{\rm Rep}(\mathbb{Z}_2)$. The generator $[2,0,0]$ is nevertheless obstructed in the matrix-factorisation analysis, see Appendix~\ref{app:even}, while the invertible line $[4,0,0]$ is not.}
This is therefore in conflict with the above expectations.

\section{Massless flows \& Matrix Factorisations}\label{sec:MF}
In order to understand the origin of this apparent inconsistency, we shall, in the following, study the above problem in the language of matrix factorisations. Again, we first need to introduce some notation before we can explain how the above problem can be phrased in this context. 

\subsection{LG description}

The $\mathcal{N}=2$ minimal models have a Landau--Ginzburg description \cite{Vafa:1988uu}. The $k^{\rm th}$ minimal model corresponds to the superpotential
\be
W (X) = X^{k+2} \ . 
\ee
The least relevant flow is induced by perturbing the superpotential via
\be\label{LGpert}
W(X) = X^{k+2} + \lambda X^{k} \ , 
\ee
and this is believed to induce the flow from the $k^{\rm th}$ minimal model to the $(k-2)^{\rm nd}$, i.e.\ this is the LG version of the above perturbation. 

The reason why this perspective is useful is that one can analyse the B-type defects of the minimal model, resp.\ the LG theory, in terms of matrix factorisations of the superpotential. More specifically, for the $k^{\rm th}$ minimal model, the topological B-type defects\footnote{The implicit assumption here is that the topological defects preserve the B-type supersymmetry, which should be true at the UV and IR fixed points. The obstruction we shall find is related to whether this should hold during the entire RG flow; we will come back to this issue in  Section~\ref{sec:BRST}.} are described by matrix factorisations of the superpotential \cite{Kapustin:2002bi,Brunner:2003dc,Kapustin:2003ga,Brunner:2007qu}
\be
W = X^d - Y^d \ , 
\ee
where $d = k+2$. Here a matrix factorisation is defined via 
\be\label{Qdef}
Q = \left( \begin{matrix} 0 & J \\ E & 0 \end{matrix} \right) \ , \qquad 
Q^2 = W \cdot  {\bf 1} \qquad (E \cdot J = J \cdot E = W \cdot {\bf 1}) \ ,
\ee
and $E$ and $J$ are, in general, matrices whose entries are polynomials in $X$ and $Y$.

Given a matrix factorisation $Q$, one can determine its bosonic and fermionic excitations in terms of the $Q$-cohomology as follows. A boson $\Phi_B$ and a fermion $\Phi_F$ are matrices of the form 
\be
\Phi_B = \left( \begin{matrix} \phi_1 & 0 \\ 0 & \phi_0 \end{matrix} \right) \ ,\qquad 
\Phi_F = \left( \begin{matrix} 0 & t_0 \\ t_1 & 0 \end{matrix} \right) \ .
\ee
The $Q$-closure condition means that 
\be
[Q,\Phi_B] = 0 = \{Q,\Phi_F\} \ , 
\ee
or in terms of components
\be\label{Qclosure}
J \phi_0 - \phi_1 J = E \phi_1 - \phi_0 E = 0 \ , \qquad \qquad
E t_0 + t_1 J = J t_1 + t_0 E=0 \ .
\ee
Furthermore, they are $Q$-exact if
\be
\Phi_B = \{Q, \tilde{\Phi}_F \} \ , \qquad \Phi_F = [Q,\tilde{\Phi}_B] \ . 
\ee

\subsection{Relation to CFT defects}

The simplest matrix factorisations of the superpotential $W_0 = X^d - Y^d$ arise from observing that 
\be\label{3.9}
W_0 = X^d - Y^d = \prod_{j=1}^{d} (X - \omega_d^j Y) \qquad \omega_d^j = e^{2 \pi i \frac{j}{d}} \ . 
\ee
For any subset $\mathcal{I} \subset \{1,\ldots,d\}$, we can then define a rank-1 matrix factorisation by
\be
E = \prod_{j\in \mathcal{I}} (X - \omega_d^{j} Y) \ , \qquad J = \prod_{j\not\in \mathcal{I}} (X - \omega_d^{j} Y) \ .
\ee
It was shown in \cite{Brunner:2005fv} that the elementary defects associated with $[L=2\ell,M=0,0]$ correspond to the simple case where $E_{[\ell]}$ consists of the factors
\be
E_{[\ell]} = \prod_{n=-\ell}^{\ell} (X - \omega_d^n Y) \ , \qquad J_{[\ell]} = \frac{W_0}{E_{[\ell]}}  \ . 
\ee

\subsection{Deformation to first order} \label{sec:1st-order}

The CFT analysis of the previous section predicts that these defects survive at least to
first order, and we can now confirm this directly. In the following we shall discuss the case when $d$ is odd, $d=2m+1$. The even case is analogous, except that the top line $[k,0,0]$ becomes an invertible simple current which survives the linear deformation.\footnote{See Appendix~\ref{app:even} for the necessary changes.} Consider the Verlinde line $[2\ell,0,0]$, whose matrix
factorisation is
\begin{equation}\label{3.12}
    E_{[\ell]} = (X-Y) \prod_{n=1}^{\ell} \bigl( X^2 - (\omega_d^n + \omega_d^{-n}) X Y + Y^2  \bigr) \ , \qquad
    J_{[\ell]} = \prod_{n=\ell+1}^{m} \bigl( X^2 - (\omega_d^n + \omega_d^{-n}) X Y + Y^2  \bigr)  \ .
\end{equation}
For the perturbed problem~\eqref{LGpert}, the corresponding defect should be a matrix
factorisation of
\be\label{Wdef}
W(\lambda) = X^d - Y^d  + \lambda \bigl(X^{d-2} - Y^{d-2} \bigr) \ ,
\ee
so we seek
\be
E_{[\ell]}(\lambda) = E_0 + \lambda E_1 + \lambda^2 E_2 + \cdots \ , \qquad
J_{[\ell]}(\lambda) = J_0 + \lambda J_1 + \lambda^2 J_2 + \cdots
\ee
with
\be
E_{[\ell]}(\lambda) \cdot J_{[\ell]}(\lambda) = W(\lambda) \ .
\ee
We will see that this can be solved at first order in $\lambda$, as the CFT analysis
predicts, but that an obstruction arises at second order that cannot be lifted, in
keeping with the puzzle of the previous section.

To see how the first-order deformation works, write
\be
Q(\lambda) = Q_0 + Q_1 \lambda + \mathcal{O}(\lambda^2) \ ,
\ee
where $Q_1$ has the form~\eqref{Qdef} with entries $E_1$ and $J_1$. At first order in
$\lambda$ the deformation problem reads
\begin{equation} \label{eq:first_order}
    E_0 J_1 + E_1 J_0 = X^{d-2} - Y^{d-2} \ ,
\end{equation}
which can be written as $\{ Q_0 , Q_1\} = X^{d-2} - Y^{d-2}$. Thus the matrix factorisation $Q_0$ can be deformed if the perturbation is $Q_0$-exact. This is indeed the case, as we shall now show. We define 
\begin{align}
E_{[\ell]}(\lambda) &= (X-Y) \prod_{n=1}^{\ell}
\Bigl( X^2 - (\omega_d^n + \omega_d^{-n}) X Y + Y^2 - \lambda \frac{(\omega_d^n - \omega_d^{-n})^2}{d}   \Bigr) \ , \label{1stsol}\\
J_{[\ell]}(\lambda) & =\prod_{n=\ell+1}^{m}
\Bigl( X^2 - (\omega_d^n + \omega_d^{-n}) X Y + Y^2 - \lambda \frac{(\omega_d^n - \omega_d^{-n})^2}{d}   \Bigr) \ , \label{1stsolJ}
\end{align}
and find by a direct calculation that their product equals 
\begin{align}
E_{[\ell]}(\lambda)  \cdot J_{[\ell]}(\lambda) & = (X-Y) \prod_{n=1}^{m}
\Bigl( X^2 - (\omega_d^n + \omega_d^{-n}) X Y + Y^2 - \lambda \frac{(\omega_d^n - \omega_d^{-n})^2}{d}   \Bigr)  \label{3.20} \\
& = X^d - Y^d  + \lambda \bigl(X^{d-2} - Y^{d-2} \bigr) + \mathcal{O}(\lambda^2) \ . \label{3.21}
\end{align}
This therefore solves the deformation problem to first order. However, the $\mathcal{O}(\lambda^2)$ term does not vanish, so this is not yet a deformation at second order. (In fact, as we are about to show, it is impossible to solve the deformation problem at second order.) As we shall see below in Section~\ref{sec:mass}, the right-hand side of (\ref{3.20}) has in fact a natural interpretation: it describes the integrable massive perturbation for which the defect can therefore be deformed. 

\subsection{Obstruction at higher order}\label{sec:massless}

In the previous section we made a specific ansatz for $E_{[\ell]}(\lambda)$ and $J_{[\ell]}(\lambda)$, see eqs.~(\ref{1stsol}) and (\ref{1stsolJ}), that demonstrated that one can deform the defect to first order, and this ansatz failed for the \emph{strictly linear} deformation
\begin{equation}\label{eq:linearW}
    W(\lambda) = X^{d} - Y^{d} + \lambda\,(X^{d-2} - Y^{d-2}) \equiv W_0 + \lambda W_1 
\end{equation}
at second order, since the $\mathcal{O}(\lambda^2)$ term in (\ref{3.21}) is nonzero. However, one may ask whether a different ansatz could also have worked to higher order. In the following we shall show that, except for the trivial matrix factorisation for which either $E$ or $J$ is equal to $1$ or $(X-Y)$, no rank-1 matrix factorisation can be deformed under the perturbation (\ref{eq:linearW}).\footnote{Here we are assuming that $d$ is odd; for $d$ even, also the top line $[k,0,0]$ is unobstructed, see below.} In fact, we shall see that for the $[2\ell,0,0]$ lines, the obstruction always arises at order $\mathcal{O}(\lambda^2)$. 

Our argument will proceed in a number of steps: first we shall show that there is an obstruction if we assume that the deformed matrix factorisation is also rank-1, see Section~\ref{sec:norank1}. Then we shall show that, in this case, the obstruction always arises at order $\mathcal{O}(\lambda^2)$, see Section~\ref{sec:order2}. Finally, we shall show that allowing the deformed matrix factorisation to have higher rank does not resolve the obstruction, see Appendix~\ref{app:rank}. For simplicity we shall continue to focus on the case where $d = 2m+1$ is odd. For even $d$ there is an extra factor $(X+Y)$ from the residual $\mathbb{Z}_2$ symmetry, and the top line $[k,0,0]$ is unobstructed. All the other lines behave as in the odd case, see Appendix~\ref{app:even} for the necessary changes.

\subsubsection{No rank-1 deformation}\label{sec:norank1} 

We begin by noting that $\lambda$ has positive scaling dimension. It is then natural to take the deformed matrix factorisation also to be polynomial in $\lambda$, so that we may work in the polynomial ring $\mathbb{C}[X,Y,\lambda]$, which is a unique factorisation domain. The argument is then purely about how $W(\lambda)$ factorises there.

As we have explained before, see eq.~(\ref{3.9}), at $\lambda=0$, the potential can be written as a product of linear factors, 
\begin{equation}\label{3.23}
    W_0 = X^d - Y^d = (X-Y)\prod_{n=1}^{m}\bigl(X^2 - (\omega_d^n+\omega_d^{-n})XY + Y^2\bigr) = \prod_{j=0}^{d-1}\bigl(X-\omega_d^{\,j}Y\bigr) \ ,
\end{equation}
and a rank-1 defect is simply a way of distributing these $d$ factors between $E$ and
$J$. (The following argument applies to arbitrary rank-1 factorisations, not just those of the form $[2\ell,0,0]$.)

However, this ceases to be the case once $\lambda$ is switched on. Indeed, the two terms $W_0$ and $W_1$ share only the linear factor
\begin{equation}
    \gcd(W_0, W_1) = X - Y \ ,
\end{equation}
since a common root $X=\omega_d^{\,n}Y$ would require $\omega_d^{\,n(d-2)}=1$ on top of
$\omega_d^{\,nd}=1$, hence $\omega_d^{2n}=1$ and so $n=0$ for odd $d$. Pulling out this
factor, let us write
\begin{equation}\label{deformed}
    W(\lambda) = (X-Y)\,\widehat{W}(\lambda) \ , \qquad
    \widehat{W}(\lambda) = \frac{W_0}{X-Y} + \lambda\,\frac{W_1}{X-Y} \ .
\end{equation}
If we assume that the deformed matrix factorisation is also rank-1, $W(\lambda)=E(\lambda)J(\lambda)$, then eq.~(\ref{deformed}) leaves little freedom: each irreducible piece, i.e.\ $(X-Y)$ and $\widehat{W}(\lambda)$, must go entirely into $E(\lambda)$ or entirely into $J(\lambda)$.\footnote{That $\widehat{W}(\lambda)$ is irreducible in $\mathbb{C}[X,Y,\lambda]$ follows from the fact that it has degree one in $\lambda$. In any factorisation one of the two factors is independent of $\lambda$, and it therefore divides both coefficients of $\widehat{W}(\lambda)$, whose greatest common divisor is $1$ by construction. The same argument applies for even $d$, see Appendix~\ref{app:even}.} However, this is only possible if the original defect, i.e.\ the one for $\lambda=0$, is trivial. This is the
matrix-factorisation counterpart of the fusion-ring obstruction of
Section~\ref{sec:deform}.

\subsubsection{Obstruction to second order}\label{sec:order2} 

The argument of the previous subsection only shows that there is an obstruction, but it does not identify at which order in $\lambda$ the obstruction appears. In order to study this question, let us consider from now on the defect $[2\ell,0,0]$, with $0<\ell<m$, i.e.\ the factorisation $W_0=E_0J_0$ for which $E_0$ is of the form
\be\label{3.26}
E_0=\prod_{|n|\le\ell}(X-\omega_d^{\,n}Y) \ . 
\ee
This is homogeneous in $(X,Y)$ of degree $2\ell+1$. We are looking for a deformation preserving the grading, i.e.\ a homogeneous function $E(\lambda)$ 
\be
E(\lambda)=E_0+\lambda E_1+\lambda^2E_2+\cdots 
\ee
of degree $2\ell+1$, with a matching $J(\lambda)$. Since $\deg\lambda=2$, each coefficient $E_p$ is itself homogeneous, of degree $2\ell+1-2p$. At first order this requires an $E_1$ of degree $2\ell-1$, with some $J_1$, satisfying
\be\label{3.28}
  E_0J_1+E_1J_0=W_1 \ ,
\ee
and at second order an $E_2$ of degree $2\ell-3$ satisfying
\be\label{3.29}
  E_0J_2+E_2J_0=-E_1J_1 \ .
\ee
This is equivalent to requiring that $E_1J_1\in(E_0,J_0)$. We want to show that no such $E_2$ exists.

For the following discussion it is convenient to pass to the affine coordinate $u=X/Y$; indeed, we can always write 
\be\label{Ydep}
E_p(X,Y)=Y^{2\ell+1-2p}E_p(u) \ , 
\ee
where $E_p(u)$ is a polynomial in $u$ of degree at most $2\ell+1-2p$. (A polynomial of higher degree would require negative powers of $Y$ and so cannot arise this way.) Thus we know that, regarded as a function of $u$, the degrees of $E_1$ and $E_2$ satisfy 
\be\label{ineq}
\deg E_1\le2\ell-1 \ , \qquad \hbox{and} \qquad \deg E_2\le2\ell-3\ .
\ee
Note that, in terms of the affine coordinate, we have from eq.~(\ref{3.26})
\be
E_0(u) = \prod_{|n|\le\ell}(u-\omega_d^{\,n})  \ . 
\ee
We also note that while (\ref{3.28}) does not have a unique solution for $E_1$ and $J_1$, 
any two solutions of (\ref{3.28}) differ by a solution of 
\be\label{E1form}
E_0(u)\,  \Delta J_1(u)+\Delta E_1(u)\, J_0(u)=0 \ .
\ee 
Since $E_0(u)$ and $J_0(u)$ have no common root, it follows that $\Delta E_1(u)$ must be of the form $\Delta E_1(u) = E_0(u) t(u)$, with $t(u)$ being a polynomial in $u$. 

A convenient particular solution to eqs.~(\ref{3.28}) and (\ref{3.29}) can be found by considering the roots of $W(u;\lambda)=0$. Since the roots of $W_0(u)$ are $u_n = \omega_d^{\,n}$ with $n=0,\ldots, d-1$, see eq.~(\ref{3.23}), we can write the roots for general $\lambda$ as\footnote{In the following we have rescaled $\lambda$ as $\lambda \mapsto \lambda Y^{-2}$.}
\be
u_n(\lambda)=\omega_d^{\,n}+\lambda\, \delta_n+\lambda^2\sigma_n+\cdots\ .
\ee
Thus the natural ansatz for $E(u)$ is 
\be\label{product}
E(u) = \prod_{|n|\le\ell} \bigl(u-u_n(\lambda)\bigr) \ , 
\ee
which is a monic factor of $W(u;\lambda)=0$ of degree $2\ell+1$. If we expand it in powers of $\lambda$, the terms of order $\lambda$ and $\lambda^2$ give particular solutions $E_1^\star(u)$ and $E_2^\star(u)$ to $E_1(u)$ and $E_2(u)$, respectively. By construction, $E^\star_1(u)$ has degree at most $2\ell$ in $u$, but actually, its degree is at most $2\ell -1$ since $\delta_n = \frac{1}{d} (\omega_d^n - \omega_d^{-n})$, and hence 
\be\label{sumdelta}
\sum_{|n|\leq \ell} \delta_n = 0 \ ,
\ee
because the roots in $E_0$ appear symmetrically. Here it is important that we are considering the specific rank-1 factorisation corresponding to $[2\ell,0,0]$. 

Because of the argument below eq.~(\ref{E1form}), any other solution for $E_1(u)$ differs from $E^\star_1(u)$ by terms of the form $E_0(u) t(u)$. However, since $E_0(u)$ has degree $2\ell+1$, $E^\star_1(u)$ is the unique solution for $E_1(u)$ with degree at most $2\ell -1$ in $u$. From now on we shall therefore fix $E_1(u)\equiv E^\star_1(u)$.

At second order, we consider the order-$\lambda^2$ term in the product (\ref{product}). This has again degree at most $2\ell$, and its coefficient of $u^{2\ell}$ equals $-\sum_{|n|\le\ell}\sigma_n$, where 
\be\label{sumsigma}
 \sum_{|n|\le\ell}\sigma_n= \frac{(d-3)(S_3-S_1)}{2d^2}\ ,
 \qquad S_p=\sum_{n=-\ell}^{\ell}\omega_d^{\,pn}\ .
\ee
This is nonzero for every nontrivial $0<\ell< m$, because
\be\label{sumsigma1}
  S_1-S_3=\frac{4\,\sin(2\ell\varphi)\,\sin\!\big((2\ell+1)\varphi\big)\,
                 \sin\!\big((2\ell+2)\varphi\big)}{\sin 3\varphi}\ ,
  \qquad \varphi=\frac{\pi}{d} \ ,
\ee
is a product of sines whose angles all lie strictly between $0$ and $\pi$. Hence
$E_2^\star(u)$ has degree exactly $2\ell$ in $u$. 

Given that $E_1(u)$ (and hence $J_1(u)$) is already uniquely fixed, any two solutions of (\ref{3.29}) differ by a solution of 
\be\label{E2form}
E_0(u)\,  \Delta J_2(u)+\Delta E_2(u)\, J_0(u)= 0 \ ,
\ee 
i.e.\ also $\Delta E_2(u)$ must be of the form $\Delta E_2(u) = E_0(u) \, t(u)$. However, since 
$E_0$ has degree $2\ell+1$, adding $E_0\, t$ can only increase the degree, and hence the minimal degree of $E_2(u)$ is $2\ell$, and the solution of that degree is $E_2^\star(u)$. This therefore does not satisfy the second inequality in (\ref{ineq}), i.e.\ it leads to negative powers of $Y$ in (\ref{Ydep}) once we reinstate both $X$ and $Y$ from the affine coordinate. Thus we have shown that among the rank-1 factorisations, the obstruction always appears at second order. Put differently, this argument shows that $E_1J_1\notin(E_0,J_0)$.

The restriction to grading-preserving deformations is in fact not needed for this conclusion. This follows from two observations. First, the choice of the first-order solution is
immaterial. Indeed, the general solution of \eqref{3.28} is $(E_1+E_0\,\mu,\ J_1-J_0\,\mu)$ with $\mu$ an arbitrary, not
necessarily homogeneous, polynomial, since the difference of two solutions satisfies $E_0\,\Delta J_1=-\Delta E_1\,J_0$, and $E_0$
shares no common factor with $J_0$. 
The terms involving $\mu$ can then be absorbed into the redefinitions $E_2\to E_2-\mu E_1$ and
$J_2\to J_2+\mu J_1-J_0\,\mu^2$, after which the second-order equation reduces to the one for the original $(E_1,J_1)$. 
Second, if the second-order equation admits any polynomial solution, it also admits a homogeneous solution with the degrees of \eqref{ineq}. To see this, note that $E_0$, $J_0$ and $E_1J_1$ are homogeneous of
degrees $2\ell+1$, $d-2\ell-1$ and $d-4$, so keeping only the terms of degree $d-4$ in the equation turns any solution into a homogeneous one of exactly these degrees. The degree count above therefore rules out every second-order deformation, homogeneous or not, and in particular any deformation organised as a formal power series in $\lambda$.

Finally, one may wonder whether the obstruction is an artefact of insisting that the deformed matrix factorisation remains rank-1. This is not the case. If the line is presented by a higher-rank matrix factorisation, obtained by adding the trivial factorisations $(1,W_0)$ and $(W_0,1)$ to the original matrix factorisation, the additional off-diagonal terms that could in principle cancel $E_1J_1$ are themselves forced to lie in the ideal $(E_0,J_0)$, so that the class $[E_1J_1]\in\mathbb{C}[X,Y]/(E_0,J_0)$ is unchanged. Since the class is also invariant under gauge equivalence, and since these two moves generate all isomorphisms in the
homotopy category of matrix factorisations \cite{Eisenbud,Yoshino}, the second-order obstruction is a property of the defect line itself, and not of the presentation chosen to represent it. The details of this argument are spelled out in Appendix~\ref{app:rank}.

\section{Massive Flows}\label{sec:massive}

The analysis of Section~\ref{sec:massless} demonstrates that the matrix factorisation corresponding to the defect $[2\ell,0,0]$ is necessarily obstructed if we work with the strictly linear deformation of eq.~(\ref{eq:linearW}).
However, we can overcome this obstruction if we are prepared to modify also the 
superpotential at higher orders in $\lambda$, and the analysis in Section~\ref{sec:1st-order}, see eqs.~(\ref{1stsol})--(\ref{3.21}) suggests a natural way to do so. In fact, as we shall explain below, by removing the obstruction order-by-order in the deformation, 
we can fix the resulting superpotential to be exactly the Chebyshev (Dickson) deformation that describes the massive flow \cite{Fendley:1991ve}.

\subsection{Massive completion to all orders}\label{sec:mass}

Let the undeformed superpotential be $W_0(X) = X^d$. As before we take $d = 2m+1$ odd, so that $m = (d-1)/2$.\footnote{For even $d = 2m$, one can additionally factor out $(X+Y)$, corresponding to the residual $\mathbb{Z}_2$ symmetry, and the rest of the analysis is essentially identical. See Appendix~\ref{app:even} for more details.} The difference of the undeformed potentials factorises into $m$ quadratic polynomials,
\begin{equation}\label{eq:facto}
    W_0(X) - W_0(Y) = (X-Y) \prod_{j=1}^{m} \left( X^2 - (\omega_d^j + \omega_d^{-j})XY + Y^2 \right) \ ,
\end{equation}
where $\omega_d = \exp(2\pi i / d)$. Recall that the matrix factorisation of the $[2\ell,0,0]$ line corresponds to the choice
\begin{align}
    E_{[\ell]}(X,Y) &= (X-Y) \prod_{n=1}^\ell \left( X^2 - (\omega_d^n + \omega_d^{-n})XY + Y^2 \right) \ , \\
    J_{[\ell]} (X,Y) &= \prod_{n=\ell+1}^m \left( X^2 - (\omega_d^n + \omega_d^{-n})XY + Y^2 \right) \ .
\end{align}

We wish to find a deformed potential $W(X, \lambda)$ by introducing a deformation parameter $\lambda$. Suppose we deform the quadratic factors by adding constant shifts $\lambda c_j$,
\begin{equation}\label{eq:deformed_ansatz}
    W(X, \lambda) - W(Y, \lambda) = (X-Y) \prod_{j=1}^{m} \left( X^2 - (\omega_d^j + \omega_d^{-j})XY + Y^2 + \lambda c_j \right) \ .
\end{equation}
Here the shifts $\lambda c_j$ are taken to be independent of $X$ and $Y$, so that the factorisation preserves all the $[2\ell,0,0]$ lines (now carrying a deformation) but not the $[2\ell,M,0]$ lines with $M \neq 0$. The crucial observation is that this ansatz is extremely rigid. Matching the linear order in $\lambda$ fixes every constant $c_j$,\footnote{Evaluating the order-$\lambda$ equation at $X=\omega_d^{n}Y$ kills every term of the sum except the $n$-th, since all the others carry the vanishing factor $X^2-(\omega_d^{n}+\omega_d^{-n})XY+Y^2$, and $c_n$ is thereby determined.} and the $c_j$ in turn fix each quadratic factor and hence --- through the product --- the full superpotential $W(X,\lambda)$ up to an additive constant. The only nontrivial question is whether constant shifts can be assembled into a genuine difference $W(X,\lambda)-W(Y,\lambda)$ to all orders in $\lambda$. We now show that they can, and that the resulting $W(X,\lambda)$ is precisely the Chebyshev (Dickson) deformation.

To this end, introduce the algebraic uniformisation $X = \xi - \frac{\lambda}{d\,\xi}$. Setting $a = -\frac{\lambda}{d}$ for notational convenience, so that $X = \xi + \frac{a}{\xi}$, we define
\begin{equation}
    W(X, \lambda) = \xi^d + \frac{a^d}{\xi^d} \ ,
\end{equation}
which is the scaled Chebyshev (equivalently, Dickson) polynomial of the first kind of degree $d$. Expanding for small $\lambda$ yields $W(X,\lambda) = X^d + \lambda X^{d-2} + \mathcal{O}(\lambda^2)$, which correctly reproduces the required linear perturbation~\eqref{LGpert}.

The factorisation theorem proved below in Appendix~\ref{app:factor} states that the difference of two such potentials factors exactly as
\begin{equation}\label{eq.4.6}
    W(X, \lambda) - W(Y, \lambda) = (X-Y) \prod_{j=1}^{m} \left( X^2 - (\omega_d^j + \omega_d^{-j})XY + Y^2 + a(\omega_d^j - \omega_d^{-j})^2 \right) \ .
\end{equation}
Recalling $a = -\frac{\lambda}{d}$, this identity reproduces the ansatz~\eqref{eq:deformed_ansatz} with
\begin{equation}
    c_j = -\frac{(\omega_d^j - \omega_d^{-j})^2}{d} \ ,
\end{equation}
and this is exactly the value already fixed at first order in Section~\ref{sec:1st-order}. The constant shifts therefore do assemble into an exact difference to all orders, so that the higher-order terms of the matrix factorisation are completely determined. By contrast, the strictly linear superpotential $X^d + \lambda X^{d-2}$ admits no such completion --- this is precisely the obstruction established in Section~\ref{sec:massless} --- and hence consistency to all orders forces the massive Chebyshev deformation within this ansatz.

\subsection{Relation to flat coordinates and integrability}

Let us discuss several broader implications of the Chebyshev-deformed superpotential. From the perspective of topological field theory, using Chebyshev polynomials rather than simple monomials as a basis for the deformation is intimately tied to the notion of flat coordinates, and hence to that of primitive forms.

The geometry of the deformation theory governed by the Chebyshev superpotential was investigated in~\cite{Dijkgraaf:1990dj,Cecotti:1991me}, where it was shown that the deformation parameter $\lambda$ is a flat coordinate of the topological Landau--Ginzburg model. Through a change of variables (the uniformisation introduced above), the Chebyshev superpotential is locally mapped to a cosine superpotential, thereby establishing a connection with the deformation theory of the $\mathcal{N}=2$ sine-Gordon model.

This correspondence shows that the Chebyshev deformation can be regarded as an integrable deformation of the $\mathcal{N}=2$ minimal models~\cite{Fendley:1991ve,Fendley:1992dm}, which fits naturally with our all-order matrix factorisations. In the $\phi_{1,3}$ deformation of the Virasoro minimal models, the choice of sign dictates whether the renormalisation group flow is massive or massless. In the $\mathcal{N}=2$ case the Chebyshev deformation unambiguously drives a \emph{massive} flow. The critical points of $W(X,\lambda)$ are the roots of $W'(X,\lambda) \propto U_{d-1}(X/(2\sqrt{a}))$, the Chebyshev polynomial of the second kind, so the infrared theory possesses $d-1$ isolated, non-degenerate massive vacua.

Furthermore, Gepner~\cite{Gepner:1990gr} showed that the chiral ring of the Landau--Ginzburg theory with a Chebyshev superpotential of degree $d$ is isomorphic to the $\mathfrak{su}(2)$ fusion ring at level $d-2$, i.e.\ at the original level $k = d-2$, which carries $d-1$ primaries. It is instructive to observe that the fusion algebra of the preserved defects extracted from our all-order matrix factorisations reproduces this (integer-spin part of the) $\mathfrak{su}(2)_{d-2}$ structure. The integer-spin subring encountered in the CFT puzzle is thereby carried intact by the massive branch of the flow.

A systematic way to compute the fusion rules of the deformed defects from the rank-1 matrix factorisations is provided by the algebraic resultant. Consider two such factorisation components,
\begin{align}
    E_{[\ell_1]}(X,Y) &= (X- Y) \prod_{j=1}^{\ell_1} \Bigl( X^2 - (\omega_d^j + \omega_d^{-j}) XY + Y^2 - \lambda\, \tfrac{(\omega_d^j - \omega_d^{-j})^2}{d} \Bigr) \ , \label{4.8} \\
    E_{[\ell_2]}(Y,Z) &= (Y- Z) \prod_{j=1}^{\ell_2} \Bigl( Y^2 - (\omega_d^j + \omega_d^{-j}) YZ + Z^2 - \lambda\, \tfrac{(\omega_d^j - \omega_d^{-j})^2}{d} \Bigr) \ . \label{4.9}
\end{align}
The fusion of the two defects corresponds to computing the resultant $R(X,Z)$ of $E_{[\ell_1]}(X,Y)$ and $E_{[\ell_2]}(Y,Z)$ with respect to the intermediate variable $Y$, see e.g.\ \cite{Brunner:2007ur}. Algebraically, this eliminates $Y$ through the determinant of the associated Sylvester matrix. We can show that the resultant factorises as
\begin{equation}\label{resultant}
    R(X,Z) = \bigl( W(X,\lambda) - W(Z,\lambda) \bigr)^{N} \prod_{\ell \,\in\, \ell_1 \otimes \ell_2} E_{[\ell]}(X,Z) \ ,
\end{equation}
for some non-negative integer $N$. The overall factor $\bigl(W(X,\lambda)-W(Z,\lambda)\bigr)^{N}$ encodes the truncation of the fusion at level $d-2$. Once it is removed, the remaining product reproduces precisely the defects appearing in the fusion channel, in agreement with the $\mathfrak{su}(2)_{d-2}$ fusion rules.\footnote{If non-simple defects were included, the decomposition of the right-hand side would no longer be unique.} A formal proof of~\eqref{resultant} follows from the standard properties of the resultant together with the divisibility $E_{[\ell]} \mid \bigl( W(X,\lambda) - W(Z,\lambda) \bigr)$, which holds for each rank-1 component of a matrix factorisation, see Appendix~\ref{app:result}.

\section{Discussion}\label{sec:discussion}

In Section~\ref{sec:MF} we have illustrated that along the massless flows between the $k^{\rm th}$ and $(k-2)^{\rm nd}$  $\mathcal{N}=2$ minimal models, induced by the superpotential deformation
\begin{equation}
    W(X)= X^{k+2} + \lambda X^k \ , 
\end{equation} 
the matrix factorisation corresponding to the defect $[2\ell,0,0]$ is obstructed at order $\lambda^2$ in the deformation. In particular, there is no consistent order-by-order deformation of the corresponding matrix factorisation if one insists on the linear deformation from above. 

One way to resolve this problem is to deform the superpotential systematically order-by-order in $\lambda$ such that the matrix factorisation describing the defects persists to all orders, see Section~\ref{sec:massive}. Microscopically, these higher-order corrections may be understood as arising from counterterms that are required to cancel the anomaly. As we explained, this leads to an integrable flow whose endpoint is a purely massive theory. However, the CFT also possesses a massless flow, whose IR endpoint is the $(k-2)^{\rm nd}$ minimal model, and we should also understand what happens in that case.

\subsection{An interpretation of the obstruction}\label{sec:5.1}
As we have explained above, the CFT analysis of Section~\ref{sec:puzzle} implies that there is an obstruction to deforming the preserved topological defect lines along the massless flow, and the matrix factorisation analysis corroborates this conclusion. The two analyses also agree in a finer respect. The obstruction is absent at first order, consistent with the lines commuting with the deformation, and it only appears at second order in $\lambda$.

While this shows that the two considerations are compatible with one another, one may still wonder what the physical reason for the obstruction is. A natural way to think about this issue is to take the formal ansatz for $E(u)$ in (\ref{product}) seriously, and continue it to large $\lambda$.\footnote{While, as we have seen, $E(u)$ does not define a legitimate matrix factorisation beyond first order (since it is not polynomial in $X$ and $Y$), we can, as we shall see momentarily, interpret it again as a conventional matrix factorisation for large $\lambda$.} Note that the IR theory only emerges in the limit $\lambda\rightarrow\infty$, which reflects that the CFT RG flow has infinite length in the Zamolodchikov metric \cite{Cvetic:1989qv}. In the limit $\lambda\rightarrow\infty$, the roots of $W(u;\lambda)$ are of the form 
\be
\hat{u}_n = \omega_{d-2}^n \ , \qquad n=0,\ldots,d-3 \ , 
\ee
together with the two ``massive'' roots for which $u \sim \pm i \sqrt{\lambda}$. In fact, if we take $\lambda$ to lie along the positive real axis along the flow, the roots rearrange themselves as 
\be\label{rootflow}
\omega_d^n  \rightarrow \left\{ \begin{array}{ll} 
\omega_{d-2}^n \qquad & \hbox{if $|n| \leq \lfloor\, \frac{d-2}{4} \rfloor$} \\[4pt]
\omega_{d-2}^{\,n-\mathrm{sgn}(n)} \qquad & \hbox{if $|n| \geq  \lceil\, \frac{d+2}{4} \rceil$}
\end{array} \right.
\ee
whereas the two roots closest to the imaginary axis, namely $\omega_d^{\pm n_\ast}$ with $n_\ast=\lfloor\, \frac{d-2}{4} \rfloor +1$, flow to the ``massive'' roots with $u \sim \pm i \sqrt{\lambda}$. 

Thus, if we ignore for the moment the issue with the middle roots (that become the ``massive'' roots in the IR), we would conclude that, in the notation of (\ref{3.12}), the RG flow induces the map 
\be
E^{\rm UV}_{[\ell]} \longrightarrow E^{\rm IR}_{[\ell]} \ , \qquad \ell \leq \left\lfloor\, \frac{d-2}{4} \right\rfloor \ , 
\ee
and this would be natural. In particular, the fusion rules of the defects for which $L_1$ and $L_2$ are small are independent of $k$, see eq.~(\ref{defectring}). However, if $E_{[\ell]}$ does not involve the middle roots (as is the case for $\ell \leq \lfloor\, \frac{d-2}{4} \rfloor$) then $J_{[\ell]}$ does, so the matrix factorisation we end up with in the IR is necessarily contaminated by the massive roots. From the CFT perspective this suggests that the resulting defect of the IR theory also acts on the massive ground states, i.e.\ it is not a pure CFT defect. 
\smallskip

\subsection{A BRST anomaly perspective}\label{sec:BRST}
We can also interpret the obstruction as an anomaly. Since the perturbing field commutes with the lines $[2\ell,0,0]$, one expects them to persist as topological defects order-by-order. The refined question is whether they persist compatibly with the supercharge. Maintaining this compatibility requires dressing the line, order-by-order in $\lambda$, with counterterms localised on it, and this dressing can fail. Thus there may be a SUSY/BRST anomaly localised on the defect.\footnote{FA thanks Davide Gaiotto for a discussion about this point.}
 
This is most clearly seen in the folded picture, where the line becomes a B-type boundary of the doubled model, and the deformed supercharge squares to the boundary superpotential $Q^{2}=W$ \cite{Warner:1995ay,Kapustin:2002bi}. The first-order boundary counterterm $\Psi_1$ that cancels the order-$\lambda$ violation is precisely the first-order factorisation $E_0J_1+E_1J_0=W_1$ already solved in Section~\ref{sec:1st-order}. However, something new can happen at second order. The supercharge has by then been corrected to $Q=Q_0+\lambda\,\Psi_1$, so squaring it generates the extra term $\Psi_1^{2}=E_1J_1$. A second-order counterterm cancelling it exists iff this term is $Q_0$-exact, and the obstruction of Section~\ref{sec:order2} is precisely that the class $[E_1J_1]$ in $\mathbb{C}[X,Y]/(E_0,J_0)$ does not vanish: this is therefore the SUSY/BRST anomaly. Since the class is unchanged by adding trivial branes or by passing to an equivalent presentation, the anomaly is a property of the line itself, not of a chosen matrix factorisation.
 
This perspective answers the puzzle of how a deformation that commutes with a line can nevertheless lead to an obstruction. Commutation is a statement about a single insertion of the perturbing field in the presence of the line, and it is exactly what guarantees the first-order dressing. The anomaly, however, sits in the collision of two insertions on the line $\Psi_1^2$, which the commutation condition does not constrain. The argument recalled in the introduction implicitly assumes that such defect-localised contact terms cause no harm. Supersymmetry makes this loophole quantitative. The contact term is computable exactly, and its class is nonzero.{\footnote{One could in principle try to dress the topological lines as in \cite{Ambrosino:2025myh,Ambrosino:2025pjj} to resolve  this anomaly.}} The same perspective also exposes the two consistent ways out: either the line fails to be realised compatibly with the supercharge, as happens along the massless flow --- whether it survives in some non-supersymmetric form acting on massive vacua as well as massless vacua remains open; or the bulk trajectory is modified so that a second-order bulk term absorbs the anomaly, which requires $[E_1J_1]=[W_2(X)-W_2(Y)]$. Remarkably, the single correction $W_2\propto X^{d-4}$ cancels the anomaly class of every line $[2\ell,0,0]$ at once, and by Section~\ref{sec:massive} the iteration of this cancellation gives rise to the Chebyshev deformation.  Demanding that the whole of $\mathfrak{so}(3)_k$ be preserved along the flow thus selects the integrable massive direction.

\subsection[The IR analysis]{\boldmath The IR analysis}
Another important perspective on the physical nature of the obstruction comes from the analysis of the deformation as it approaches the IR. To this end, it is instructive to compare our result to the analysis of the bosonic minimal models. In this case, the analogous flow $\mathcal{M}_{p}\to\mathcal{M}_{p-1}$ is driven by
$\phi_{1,3}$ in the ultraviolet and approached by the irrelevant operator $\phi_{3,1}$,
of dimension $2+\tfrac{4}{p-1}$, in the infrared. Both are spinless primaries, and both
are left invariant by the topological lines that survive the flow, so the same symmetry
is visible at both endpoints of the RG flow. For the $\mathcal{N}=2$ flow studied here the analogous infrared operator behaves differently, and it is worth spelling this out.

Recall from Section~\ref{sec:deform} that the ultraviolet perturbation $(k,k,2)$ commutes
with the integer-spin lines $[L,0,0]$ with $L$ even, and that the bulk flow ends at the $(k-2)^{\rm nd}$ model. In order to understand the flow from the IR perspective, we now write $X=\lambda^{-1/k}x$, 
\be
W=X^{d}+\lambda X^{d-2} \sim x^{k}+\lambda^{-(k+2)/k}x^{k+2} \ ,
\ee
where $d=k+2$. Thus the IR theory is approached by the operator $x^{k+2}$. By the $\mathcal{N}=2$
non-renormalisation theorem the superpotential is not corrected, and indeed
$x^{k+2}=\tfrac1k\,x^{3}W_{\rm IR}'$ lies in the ideal $(W_{\rm IR}')$, so that it is a
trivial deformation of the \emph{chiral ring}. It is, however, not redundant. The
redefinition $x\to x-\tfrac{g}{k}x^{3}$ removes it from the superpotential only at the
cost of generating a K\"ahler term, 
\be
\delta K=-\tfrac{g}{k}\,(x^{3}\bar{x}+\mathrm{c.c.})\ ,
\ee
and thus the infrared approach is governed by the K\"ahler deformation
$\int d^4\theta\,(x^{3}\bar{x}+\mathrm{c.c.})$.

With respect to the $\mathfrak{su}(2)$ symmetry, the integrand decomposes as
\be
x^{3}\,\bar x\sim(3,3,0)\times(1,-1,0)\ , 
\ee
with $3\otimes1=2\oplus4$, and with $\mathfrak{u}(1)$ label $3-1=2$. The leading piece $(2,2,0)=x^{2}$ is a chiral primary. Since $x^{2}=\Phi^{2}$ is a chiral superfield, $\int d^4\theta\,(2,2,0)=\int d^4\theta\,x^{2}$ is a total
derivative and does not enter the action. The deformation is therefore carried by the
non-BPS piece $(4,2,0)$ (its conjugate $\bar x^{3}x$ supplying $(4,-2,0)$), with
$h_{L}=h_{R}=\tfrac5k$, whose full-superspace descendant does not vanish and has dimension
\begin{equation}
    \Delta = 2h+2 = 2+\tfrac{10}{k}\ .
\end{equation}
This is therefore irrelevant, as expected.\footnote{The discussion should be modified for $d\le 7$ due to the absence of $(4,2,0)$. This is similar to the case of the bosonic minimal model flow with $p=4$, see \cite{Nakayama:2024msv}.}

What changes is the fate of the symmetry. Since $[L,0,0]$ is a pure $\mathfrak{su}(2)$ line,
it only leaves a bulk scalar of label $l$ invariant provided that
\begin{equation}\label{eq:commute-ir}
    \frac{\sin\!\big(\tfrac{\pi(L+1)(l+1)}{k}\big)}{\sin\!\big(\tfrac{\pi(l+1)}{k}\big)}
    =
    \frac{\sin\!\big(\tfrac{\pi(L+1)}{k}\big)}{\sin\!\big(\tfrac{\pi}{k}\big)}\ .
\end{equation}
For nonzero $l$ the equality holds in two cases. If $l=k-2$ the two sides differ by
$(-1)^{L}$, so that it holds for every even $L$. If instead $L=k-2$ they differ by
$(-1)^{l}$, so that it holds for every even $l$. The ultraviolet perturbation is the
simple current of the level-$k$ theory, and it obeys the corresponding integer-spin
selection rule. The infrared operator, by contrast, has $l=4$. For odd $d$ the label
$k-2$ is odd, so neither case applies to a nonzero even $L$, and
\eqref{eq:commute-ir} fails for all of them. For even $d$ the only exception is
$L=k-2$, which is the infrared image of the invertible line $[k,0,0]$ that
Appendix~\ref{app:even} finds to be unobstructed, so that the two analyses agree there
as well.\footnote{In the single case $k=6$ one has $l=4=k-2$,
so that every even $L$ is exempt. The infrared criterion is then vacuous, and in particular it does
not detect the second-order obstruction that the matrix-factorisation analysis finds for the ultraviolet
lines $[2,0,0]$ and $[4,0,0]$, see Appendix~\ref{app:even}.}
The lines preserved by the ultraviolet perturbation are therefore not preserved by the
operator that governs the infrared approach. 
Thus we can see the obstruction also from the IR perspective: the infrared operator
is a $D$-term descendant of a \emph{non-BPS} field, and it does not commute with the would-be surviving lines.

\subsection{Comparison with the RG-defect approach}

A complementary, non-perturbative perspective on the same question was developed
in \cite{Brunner:2020lrm}, where the flow between two minimal models was
implemented by an RG defect $\mathcal{R}$
\cite{Brunner:2007ur,Klos:2019axh,Gaiotto:2012np}. In this approach a topological defect $\mathcal{T}_{\rm UV}$ survives if it possesses an infrared image, $\mathcal{T}_{\rm IR}$, such that 
\be\label{cond1}
\mathcal{T}_{\rm IR}=\mathcal{R}*\mathcal{T}_{\rm UV}*\mathcal{R}^{\dagger} \ .
\ee
One can also require the stronger condition that $\mathcal{R}$ is transparent to the defect $\mathcal{T}_{\rm UV}$, i.e.\ that
\be\label{cond1s}
\mathcal{T}_{\rm IR}*\mathcal{R}=\mathcal{R}*\mathcal{T}_{\rm UV}\ .
\ee
Unless a subgroup of the invertible $\mathbb{Z}_d$ phase symmetry $X\to\omega_d X$ is
preserved along the flow, only the identity, a symmetry defect related to
spectral flow, and a bound state of the two were found to possess an infrared image, i.e.\ satisfy (\ref{cond1}). Furthermore, the stronger condition (\ref{cond1s}) was only satisfied by the identity defect. The mechanism eliminating all other defects is that some supersymmetric vacua become massive and decouple \cite{Gaberdiel:2007gv}. The corresponding critical points of the superpotential run off to infinity in field space, together with the branes attached to them. For our flow these are the two critical points at $X\sim\pm i\sqrt{\lambda}$, whose
counterparts in our description are the two roots of the factorised superpotential
difference that run off to infinity, see Section~\ref{sec:5.1}. For odd $d$ the
perturbation $\lambda X^{d-2}$ breaks the $\mathbb{Z}_d$ phase symmetry
completely, so our flow does not respect this phase symmetry. In particular, the lines $[2\ell,0,0]$
with $0<2\ell<k$ are then not preserved, in agreement with our analysis. On the other hand, for even $d$, the subgroup $\mathbb{Z}_2\subset\mathbb{Z}_d$ generated by $X\to-X$ survives the
perturbation, and thus the invertible top line $[k,0,0]$ of Appendix~\ref{app:even} is preserved. 

While these predictions agree with what we have found, our analysis is finer in that it identifies that the failure happens at order $\lambda^2$, and that it takes the form of a defect-localised anomaly, see Section~\ref{sec:BRST}. The first order is protected by the commutation of the perturbing field with the lines, so that no counterexample can arise at leading order. Moreover, following the deformation order-by-order shows what it would take to avoid the failure. Cancelling the obstruction at every order forces the bulk trajectory onto the Chebyshev deformation, along which the entire family of $\mathfrak{so}(3)_k$ defects survives together with their fusion ring (Section~\ref{sec:massive}). From the perspective of the RG defect, the massless and the Chebyshev trajectories simply end in different theories, i.e.\ correspond to different RG defects. On the other hand, the perturbative analysis exhibits them as the two resolutions of one and the same second-order anomaly.

\subsection{Comparison with the parafermion flow}

It is also instructive to compare our findings to the flows of the $\mathbb{Z}_k$ parafermion theory $\mathcal{P}_k=\mathfrak{su}(2)_k/\mathfrak{u}(1)_k$ that is built on the same $\mathfrak{su}(2)_k$ algebras as the $\mathcal{N}=2$ minimal models, and for which 
the same integer-spin category $\mathfrak{so}(3)_k\cong\mathfrak{su}(2)_k^{\,\mathrm{even}}$,
generated by the lines $[L,0]$ with $L$ even, acts as a set of topological
defects \cite{Benedetti:2026drn,Zhang:2026gqp}. Its least relevant
deformation is by the $\mathbb{Z}_k$-charged primary $\mathcal{O}$ of weight
$h=1-\frac{1}{k}$ together with its conjugate,
\be
\delta S=\int d^2x\,(\lambda{\mathcal O}+\lambda^*{\mathcal O}^{*})\ , 
\ee
an $\mathfrak{su}(2)_k$ singlet that is transparent with respect to the lines $[L,0]$. The
category $\mathfrak{so}(3)_k$ is therefore preserved along the entire flow. Depending on
the phase of this complex coupling, the flow is either massless and ends at the
unitary minimal model $\mathcal{M}_{k+1}$ \cite{Fateev:1991bv,Benedetti:2026drn,Zhang:2026gqp},
or massive and gaps the theory as in the associated ${\rm O}(3)$ sigma
model \cite{Fateev:1991bv}, 
see also a related massless integrable RG flow studied in \cite{Ahn:2024sxt}.

The main difference to the $\mathcal{N}=2$ flows concerns the massless endpoint, which takes the form 
\be
\mathcal{M}_{k+1}= \frac{\mathfrak{su}(2)_{k-1}\times\mathfrak{su}(2)_1}{\mathfrak{su}(2)_k} 
\ee
in the parafermionic case. In particular, it contains $\mathfrak{su}(2)_k$ in its coset description, and contains defects that generate $\mathfrak{so}(3)_k$ in the infrared \cite{Benedetti:2026drn}. As a consequence it does not exhibit the problems we have encountered above for the $\mathcal{N}=2$ minimal models.

\subsection{Integrability and defect lines}

Finally, we briefly discuss the relation to integrability. The $\mathcal{N}=2$ Landau--Ginzburg model with the Chebyshev deformation is integrable \cite{Fendley:1991ve}, with a soliton 
$S$-matrix compatible with the preserved topological defect lines of $\mathfrak{so}(3)_k$. The $S$-matrix factorises into an $A_{k+1}$ RSOS factor --- which carries precisely the $\mathfrak{su}(2)_k$ structure the lines act on --- and a basic $\mathcal{N}=2$ factor. The lines commute with the scattering up to the similarity transformation by quantum dimensions that any non-invertible symmetry requires on multi-particle states \cite{Copetti:2024dcz}. In fact, we were led to the Chebyshev deformation by the condition that all the $\mathfrak{so}(3)_k$ defect lines are preserved, and thus integrability may be a consequence of the requirement that a large number of topological defect lines are preserved.

\subsection*{Acknowledgements}  
We thank Christian Copetti, Yichul Choi, Davide Gaiotto, Andrei Katsevich, Igor Klebanov, Lasse Merkens, Brandon Rayhaun, Shu-Heng Shao, Zimo Sun, and Yifan Wang for discussions.
Research at Perimeter Institute is supported in part by the
Government of Canada through the Department of Innovation, Science and Economic Development Canada and
by the Province of Ontario through the Ministry of Colleges and Universities. MRG is grateful to the Yukawa Institute at Kyoto University for hospitality during an early stage of this work, and IAS Princeton, where he was supported by a grant from the Adler Family Fund, during a later stage. 
The work of MRG is supported by the Simons Foundation grant
994306 (Simons Collaboration on Confinement and QCD Strings), as well as the
NCCR SwissMAP that is funded by the Swiss National Science Foundation.
YN is supported in part by JSPS KAKENHI Grant Number 21K03581 and 26K00699. 

\appendix

\section{Higher-rank deformations}\label{app:rank}

In Section~\ref{sec:massless} we only considered matrix factorisations that remain rank-1, even after the deformation. However, in general we may think of the rank-1 matrix factorisation as being part of a higher-rank matrix factorisation, for which the other factors are ``trivial'', and one may wonder whether this may allow one to overcome the obstruction of Section~\ref{sec:order2}. In the following we shall show that this is not the case. 

Let us thus begin with a higher-rank matrix factorisation that consists of the matrix factorisation corresponding to $[2\ell,0,0]$, together with copies of the trivial factorisations of $W_0$, i.e.\ the pairs $(1,W_0)$ and $(W_0,1)$. The corresponding matrices $E(\lambda)$ and $J(\lambda)$ can then be written in block-diagonal form,  and they obey
\be\label{matrixa}
  E(\lambda)\,J(\lambda)=W(\lambda)\,\mathbf{1}\ ,
\ee
where, at $\lambda=0$, the original rank-1 matrix factorisation appears in the $(1,1)$ position, while the trivial factorisations make up the remaining diagonal entries.  At zeroth order in $\lambda$ --- we shall use superscripts to keep track of the order in $\lambda$ --- both $E^{(0)}$ and $J^{(0)}$ are diagonal, with 
$E^{(0)}_{11}=E_0$ and $J^{(0)}_{11}=J_0$, while each further diagonal entry is equal to either
$(E^{(0)}_{rr},J^{(0)}_{rr})=(1,W_0)$ or $(W_0,1)$, where $r>1$.

Next we expand $EJ=W\mathbf{1}$ order-by-order in $\lambda$. Then the $(i,j)$ entry at order $\lambda^n$ equals 
\be
\sum_{a+b=n}\sum_l E^{(a)}_{il}J^{(b)}_{lj}\ .
\ee
This has to equal $W \cdot {\bf 1}$. Focusing on the $(1,1)$ entry, the first-order (in $\lambda$) condition is 
\be
E_0J_1+E_1J_0=W_1 \ , 
\ee
which agrees with the previous rank-1 analysis. Thus $E_1$ and $J_1$ are the solutions found in Section~\ref{sec:order2}, and $E^{(1)}_{11}J^{(1)}_{11}=E_1J_1$ is the rank-1 obstruction. At second order in $\lambda$, the $(1,1)$ entry gives rise to the equation 
\be\label{cond2}
  E_0J^{(2)}_{11}+E^{(2)}_{11}J_0
  +\underbrace{E^{(1)}_{11}J^{(1)}_{11}}_{=\,E_1J_1}
  +\sum_{r\ge2}E^{(1)}_{1r}J^{(1)}_{r1}=0 \  .
\ee
Thus we may potentially lift the rank-1 obstruction $E_1 J_1$ by the last sum (the sum over $r\geq 2$).

However, we are not free to choose these off-diagonal entries $E^{(1)}_{1r}$ and $J^{(1)}_{r1}$ arbitrarily since $W\cdot \mathbf{1}$ is diagonal, and hence the $(1,r)$ and $(r,1)$ entries of (\ref{matrixa}) give rise to the conditions (at first order in $\lambda$)
\be\label{3.44}
  E_0J^{(1)}_{1r}+E^{(1)}_{1r}J^{(0)}_{rr}=0 
  \qquad \hbox{and} \qquad 
  E^{(0)}_{rr}J^{(1)}_{r1}+E^{(1)}_{r1}J_0=0 \ ,
\ee
where $r>1$ is fixed (i.e.\ no sum). By assumption, $E^{(0)}_{rr}=1$ or $E^{(0)}_{rr}=W_0$, so we need to discuss the two cases. If $E^{(0)}_{rr}=1$, then the second equation of (\ref{3.44}) gives
\be
J^{(1)}_{r1}=-E^{(1)}_{r1}\, J_0\ , 
\ee
and hence the term $E^{(1)}_{1r}J^{(1)}_{r1}$ in (\ref{cond2}) is a multiple of $J_0$ and thus can be absorbed into redefining $E^{(2)}_{11}$. On the other hand, if $E^{(0)}_{rr}=W_0$, then $J^{(0)}_{rr}=1$, and the first equation of eq.~(\ref{3.44}) leads to 
\be
E^{(1)}_{1r}=-E_0J^{(1)}_{1r} \ , 
\ee
and hence the term $E^{(1)}_{1r}J^{(1)}_{r1}$ is a multiple of $E_0$ (which can thus be absorbed into redefining $J^{(2)}_{11}$). Thus, after these redefinitions, we have reduced the problem to the original obstruction problem, and it follows that the off-diagonal correction terms in (\ref{cond2}) cannot lift the obstruction. 

The computation above concerns the particular higher-rank presentation built
from trivial factorisations. More invariantly, the conclusion we just reached is an
instance of a general fact: the second-order obstruction depends only on the
defect line, not on the particular matrix factorisation used to represent it.
Two matrix factorisations $(E,J)$ describe the same line precisely when they are
related by a finite sequence of two moves: a change of basis $E\mapsto gEh^{-1}$,
$J\mapsto hJg^{-1}$ with $g,h\in GL_N(R)$ and $R=\mathbb{C}[X,Y]$ (gauge
equivalence); and the addition or removal of the trivial factorisations
$(1,W_0)$, $(W_0,1)$ (stabilisation). Equivalently, these two moves generate all
isomorphisms in the homotopy category of matrix factorisations
\cite{Eisenbud,Yoshino}.\footnote{Two matrix factorisations are related by (gauge) and (stabilisation) iff their cokernels are isomorphic up to free summands \cite{Eisenbud,Yoshino}.} For the obstruction to be a property of the line, rather
than of the chosen representative, it must be invariant under both.

Invariance under stabilisation is precisely what the computation above
establishes. The off-diagonal corrections $E^{(1)}_{1r}J^{(1)}_{r1}$ in
(\ref{cond2}) are forced by (\ref{3.44}) to be multiples of $E_0$ or of $J_0$;
they therefore lie in the ideal $(E_0,J_0)$ and can be reabsorbed into
$E^{(2)}_{11}$ and $J^{(2)}_{11}$ without altering the class $[E_1J_1]$ in
$R/(E_0,J_0)$. Invariance under gauge equivalence is immediate: conjugating a
deformation of $(E,J)$ by the ($\lambda$-dependent) matrices $g,h$ produces a
deformation of $(gEh^{-1},hJg^{-1})$ order-by-order, so a second-order lift of
one presentation exists if and only if it does for the other. 
Since the two moves generate all isomorphisms, the non-vanishing of $[E_1J_1]$
established for the rank-1 presentation persists for \emph{every} presentation of
the same line, of arbitrary rank. The second-order obstruction is thus an
invariant of the defect line itself, and not of the matrix factorisation chosen
to represent it.

\section{\texorpdfstring{Proof of the factorisation theorem eq.~(\ref{eq:facto})}{Proof of the factorisation theorem}}\label{app:factor}

Let $d = 2m+1$ be an odd integer and $\omega_d = \exp(2\pi i / d)$. We uniformise both variables and write $X = \xi + a/\xi$ and $Y = \eta + a/\eta$, in the notation of Section~\ref{sec:mass}, see also Appendix~\ref{app:result}. The difference of the superpotentials is then
\begin{equation}
    W(X, \lambda) - W(Y, \lambda) = \left( \xi^d + \frac{a^d}{\xi^d} \right) - \left( \eta^d + \frac{a^d}{\eta^d} \right) = (\xi^d - \eta^d)\left( 1 - \frac{a^d}{\xi^d \eta^d} \right) \ .
\end{equation}
Factoring each term with the roots of unity $\omega_d^s$ ($s = 0, \dots, d-1$) and recombining the factors carrying the same index $s$, we find
\begin{align}
    W(X, \lambda) - W(Y, \lambda) &= \prod_{s=0}^{d-1} \left( \xi - \omega_d^s \eta \right)\left( 1 - \frac{a}{\omega_d^s\, \xi \eta} \right) \nonumber \\
    &= \prod_{s=0}^{d-1} \left( X - \Bigl( \omega_d^s \eta + \omega_d^{-s} \frac{a}{\eta} \Bigr) \right) \ ,
\end{align}
where we have used $\xi + a/\xi = X$. We now group the factors into the symmetric pairs $s = j$ and $s = d-j \equiv -j$ for $j = 1,\dots,m$, together with the singlet $s = 0$. The $s=0$ factor is simply $X - (\eta + a/\eta) = X - Y$. For each conjugate pair $j$,
\begin{align}
    F_j &= \left[ X - \Bigl( \omega_d^j \eta + \omega_d^{-j} \frac{a}{\eta} \Bigr) \right]\left[ X - \Bigl( \omega_d^{-j} \eta + \omega_d^{j} \frac{a}{\eta} \Bigr) \right] \nonumber \\
    &= X^2 - (\omega_d^j + \omega_d^{-j})\, X \Bigl( \eta + \frac{a}{\eta} \Bigr) + \left( \eta^2 + \frac{a^2}{\eta^2} + a(\omega_d^{2j} + \omega_d^{-2j}) \right) \ . \label{B.3}
\end{align}
Substituting $Y = \eta + a/\eta$ and using $\eta^2 + a^2/\eta^2 = Y^2 - 2a$, the constant term becomes
\begin{equation}
    Y^2 + a\bigl(\omega_d^{2j} + \omega_d^{-2j} - 2\bigr) = Y^2 + a(\omega_d^j - \omega_d^{-j})^2 \ .
\end{equation}
Hence $F_j = X^2 - (\omega_d^j + \omega_d^{-j}) XY + Y^2 + a(\omega_d^j - \omega_d^{-j})^2$, and collecting the singlet together with the $m$ conjugate pairs yields
\begin{equation}
    W(X, \lambda) - W(Y, \lambda) = (X-Y) \prod_{j=1}^{m} \left( X^2 - (\omega_d^j + \omega_d^{-j})XY + Y^2 + a(\omega_d^j - \omega_d^{-j})^2 \right) \ ,
\end{equation}
which completes the proof. \hfill$\square$

\section{\texorpdfstring{Proof of factorisation of resultant, eq.~(\ref{resultant})}{Proof of the resultant}}\label{app:result}

It is convenient to uniformise the Chebyshev variables. With $a=-\lambda/d$ write
\be
  X=\xi+\tfrac a\xi\ ,\qquad Y=\eta+\tfrac a\eta\ ,\qquad Z=\zeta+\tfrac a\zeta\ ,
\ee
so that $W(X,\lambda)=\xi^{d}+a^{d}/\xi^{d}$ is invariant under $\xi\mapsto a/\xi$. Writing
\be\label{gdef}
  g_r(X,Y):=X^{2}-(\omega_d^{r}+\omega_d^{-r})XY+Y^{2}-\lambda\tfrac{(\omega_d^{r}-\omega_d^{-r})^{2}}{d} \ ,   \quad(r=1,\dots,m)
\ee
for the quadratic blocks, the factorisation theorem, see eq.~(\ref{eq:facto}), reads
\be
W(X,\lambda)-W(Y,\lambda)=(X-Y)\prod_{r=1}^{m}g_r(X,Y) \ , 
\ee
and the expression in (\ref{4.8}) becomes
\begin{equation}\label{ELdef}
  E_{[\ell]}(X,Y)=(X-Y)\prod_{r=1}^{\ell}g_r(X,Y)\ ,\qquad
  E_{[m]}(X,Y)=W(X,\lambda)-W(Y,\lambda)\ ;
\end{equation}
in particular $E_{[\ell]}\mid\big(W(X,\lambda)-W(Y,\lambda)\big)$, with cofactor $J_{[\ell]}=\prod_{r=\ell+1}^{m}g_r$.
Here we have spelled out the odd case $d=2m+1$. For even $d=2m$ the middle block degenerates to
$g_{m}=(X+Y)^{2}$, and the factorisation theorem reads instead
\be\label{deven}
W(X,\lambda)-W(Y,\lambda)=(X-Y)(X+Y)\prod_{r=1}^{m-1}g_r(X,Y)\ ,
\ee
see Appendix~\ref{app:even}, where now 
$E_{[\ell]}=(X-Y)\prod_{r=1}^{\ell}g_r$ has cofactor $J_{[\ell]}=(X+Y)\prod_{r=\ell+1}^{m-1}g_r$, so that
$E_{[m]}=W(X,\lambda)-W(Y,\lambda)$ holds for odd $d$ only. The argument below is written so as to cover both cases. We use two elementary facts:
\begin{itemize}
\item[(i)] for $X'=\mu+a/\mu$, \ $X-X'=(\xi-\mu)\,\dfrac{\xi\mu-a}{\xi\mu}\ ;$
\item[(ii)] $g_r$ depends on $r$ only through $\omega_d^{r}+\omega_d^{-r}$ and $(\omega_d^{r}-\omega_d^{-r})^{2}$,
  hence $g_r=g_{-r}=g_{d-r}$.
\end{itemize}
 Below we shall also need $E_{[\ell]}$ for $\ell>m$, see in particular eq.~(\ref{Rnaive}), where $\ell$ runs up to $\ell_1+\ell_2$, and eq.~(\ref{reflection}). For that we  take the first equation of (\ref{ELdef}) as the
definition, with $g_r$ being defined by eq.~(\ref{gdef}) for every $r\in\mathbb{Z}$. Note that by (ii), $g_r$ only depends on $r$ modulo $d$ and up to $r\mapsto-r$. We should mention that for $\ell>m$ the resulting $E_{[\ell]}$ no
longer divides $W(X,\lambda)-W(Y,\lambda)$.

\paragraph{Resultant as a product of root differences.}
As a polynomial in $Y$, $E_{[\ell_1]}(X,Y)$ has degree $2\ell_1+1$, leading coefficient $-1$, and roots $Y_s=\omega_d^{s}\xi+\omega_d^{-s}a/\xi$ ($s=-\ell_1,\dots,\ell_1$), where we have used that $g_r(X,Y)$ factorises as in eq.~(\ref{B.3}). Similarly, we can write 
\be
E_{[\ell_2]}(Y,Z)=\prod_{t=-\ell_2}^{\ell_2}(Y-Z_t) \ , \qquad Z_t=\omega_d^{t}\zeta+\omega_d^{-t}\frac{a}{\zeta} \ . 
\ee
The product formula for the
resultant in the roots of its first argument gives, up to the overall sign fixed by conventions,
\begin{equation}\label{Rprod}
  R(X,Z)=\operatorname{Res}_Y\!\big(E_{[\ell_1]}(X,Y),E_{[\ell_2]}(Y,Z)\big)
  \;\doteq\;\prod_{s=-\ell_1}^{\ell_1}\prod_{t=-\ell_2}^{\ell_2}\big(Y_s-Z_t\big)\ .
\end{equation}

\paragraph{Collecting the multiplicities.} Using (i) from above, we
can rewrite the factors as
\be\label{YZfac}
Y_s-Z_t=(\omega_d^{s}\xi-\omega_d^{t}\zeta)\,\dfrac{\omega_d^{s+t}\xi\zeta-a}{\omega_d^{s+t}\xi\zeta}
=\omega_d^{-t}\,(\xi-\omega_d^{\,t-s}\zeta)\,\dfrac{\omega_d^{\,s+t}\xi\zeta-a}{\xi\zeta}\ .
\ee
Inserting this into \eqref{Rprod}, the double product splits into factors that depend on $(s,t)$ only
through $r:=t-s$, and factors that depend on it only through $q:=s+t$. Since $\sum_t t=0$, the overall $\omega_d^{-t}$ phase cancels, and we are left with
\be\label{Rsplit}
 R(X,Z)\;\doteq\;\frac{1}{(\xi\zeta)^{(2\ell_1+1)(2\ell_2+1)}}\,
 \prod_{r}\bigl(\xi-\omega_d^{\,r}\zeta\bigr)^{N_r}\,
 \prod_{q}\bigl(\omega_d^{\,q}\xi\zeta-a\bigr)^{M_q}\ ,
\ee
where $N_r=\#\{(s,t):t-s=r\}$ and $M_q=\#\{(s,t):s+t=q\}$, and $r$ and $q$ run over the values 
\be
r,  q \in\bigl\{- (\ell_1+\ell_2), - (\ell_1 + \ell_2) +1, \ldots, (\ell_1+\ell_2)-1\ , \ (\ell_1+\ell_2) \bigr\}\ . 
\ee
Since $t\mapsto-t$ is a bijection of $\{-\ell_2,\dots,\ell_2\}$ onto itself under which 
$s+t\mapsto-(t-s)$, it follows that 
\be\label{MN}
 M_q=N_{-q} = N_q \ ,
\ee
where we have used that $N_{-q}=N_q$ as a consequence of the bijection $(s,t)\mapsto(-s,-t)$. Thus we can replace $q\mapsto r$ in the second product of \eqref{Rsplit}, and hence write 
\be\label{Rrelabel}
R(X,Z)\;\doteq\;\frac{1}{(\xi\zeta)^{(2\ell_1+1)(2\ell_2+1)}}\,
 \prod_{r=-(\ell_1+\ell_2)}^{\ell_1+\ell_2} 
 \Bigl( \bigl(\xi-\omega_d^{\,r}\zeta\bigr)\,  \bigl(\omega_d^{\,r}\xi\zeta-a\bigr)\Bigr)^{N_r}\ ,
 \ee
where
\be\label{Nr}
 N_r=\#\{\ell:|\ell_1-\ell_2|\le \ell\le\ell_1+\ell_2,\ \ell\ge|r|\}
\ee
are the weight multiplicities of the $\mathfrak{su}(2)$ tensor product.
The term with $r=0$ equals, using again (i) from above,
\be
(\xi-\zeta)(\xi\zeta-a)=\xi\zeta\,(X-Z) \ , 
\ee
while for $r\neq 0$, we combine the terms with $\pm r$ to find\footnote{To verify \eqref{block}, set $u=\xi\zeta$,
$P=\xi^{2}+\zeta^{2}$, $Q=u^{2}+a^{2}$ and
$c_r=\omega_d^{r}+\omega_d^{-r}$. Then $X^{2}+Z^{2}=PQ/u^{2}+4a$ and
$XZ=(Q+aP)/u$, so that
$g_r=X^{2}-c_rXZ+Z^{2}+a(c_r^{2}-4)=\bigl(P-c_ru\bigr)\bigl(Q-a c_r
u\bigr)/u^{2}$, which is the
left-hand side of \eqref{block} divided by $u^{2}$.}
\be\label{block}
 (\xi-\omega_d^{r}\zeta)(\xi-\omega_d^{-r}\zeta)\,(\omega_d^{r}\xi\zeta-a)(\omega_d^{-r}\xi\zeta-a)
 =(\xi\zeta)^{2}\,g_r(X,Z)\ .
\ee
Since $N_0+2\sum_{r\ge1}N_r=(2\ell_1+1)(2\ell_2+1)$, all powers of
$\xi\zeta$ cancel against the prefactor in \eqref{Rrelabel}, and we find
\begin{equation}\label{Rnaive}
 R(X,Z)\;\doteq\;(X-Z)^{N_0}\prod_{r=1}^{\ell_1+\ell_2}g_r(X,Z)^{N_r}
 \;=\;\prod_{\ell=|\ell_1-\ell_2|}^{\ell_1+\ell_2}E_{[\ell]}(X,Z)\ .
\end{equation}
Here the last equality holds because $\#\{\ell\in[|\ell_1-\ell_2|,\ell_1+\ell_2]:\ell\ge r\}=N_r$, see
(\ref{Nr}). This is the \emph{naive} (Clebsch--Gordan) fusion product.

\paragraph{Level truncation.}
Because of (ii) above, i.e.\ $g_r=g_{-r}=g_{d-r}$,  it follows from \eqref{ELdef} that 
\begin{align}
E_{[\ell]}(X,Z)\,E_{[d-1-\ell]}(X,Z) & =(X-Z)^2 \prod_{r=1}^{\ell}g_r(X,Z) \, \prod_{r=1}^{d-1-\ell}g_r(X,Z) \nonumber \\
& = (X-Z)^2 \prod_{r=1}^{d-1}g_r(X,Z)   = (X-Z)^2 \prod_{r=1}^{m}g_r(X,Z)^2 \nonumber \\
& =  \big(W(X,\lambda)-W(Z,\lambda)\big)^{2}\ ,\label{reflection}
\end{align}
where $0\le\ell\le d-1$, and we have assumed that $d=2m+1$ is odd. For even $d$, the middle block degenerates to $g_{m}=(X+Z)^{2}$, and (\ref{reflection}) remains true because of eq.~(\ref{deven}).
If we define $\ell_{\max}$ via 
\be
\ell_{\max} =\min\bigl(\ell_1+\ell_2,\ d-2-\ell_1-\ell_2\bigr)\ ,
\ee
the reflection $\ell\mapsto d-1-\ell$ maps the set of indices above the bound, $\{\ell:\ell_{\max}<\ell\le\ell_1+\ell_2\}$, to itself.
The factors of \eqref{Rnaive} with $\ell>\ell_{\max}$ therefore recombine through \eqref{reflection} into
powers of $W(X,\lambda)-W(Z,\lambda)$: for odd $d$ they pair up as $\{\ell,d-1-\ell\}$, together with a single
factor $E_{[m]}=W(X,\lambda)-W(Z,\lambda)$ at the fixed point $\ell=m=\frac{1}{2}(d-1)$, whereas for even $d$ the
reflection acts freely and they pair up completely. The factors with $\ell\le\ell_{\max}$ are left
untouched. Thus, altogether, we end up with 
\begin{equation}
  R(X,Z)\;\doteq\;\big(W(X,\lambda)-W(Z,\lambda)\big)^{N}\!\!\prod_{\ell\in\ell_1\otimes\ell_2}\!\!E_{[\ell]}(X,Z)\ ,
  \qquad N=N_0-|\ell_1\otimes\ell_2|\ \ge 0\ ,
\end{equation}
where $\ell_1\otimes\ell_2=\{\ell:|\ell_1-\ell_2|\le \ell\le \ell_{\max}\}$ is the level-$(d-2)$ truncated fusion, and 
$N$ counts the channels removed by truncation. \hfill$\square$

\section{\texorpdfstring{The even case \boldmath$d=2m$}{The even case d=2m}}\label{app:even}

Throughout the main text we assumed $d=2m+1$ odd. Here we collect the
modifications for $d=2m$ even. The single new ingredient is the residual
$\mathbb{Z}_2$ symmetry $X\to -X$ of $W_0=X^{d}$ (which is only a symmetry provided that $d$ is
even).\footnote{$X \to -X$ is still a symmetry of the theory when $d$ is odd if we accompany it with a chiral fermion number transformation, leading to an $R$-symmetry that does not commute with the supercharge. It is outside of our scope to discuss symmetries that do not commute with the supercharge.} It has two consequences: a second linear factor $(X+Y)$ in $W_0$, and the
fact that the top integer-spin line $[k,0,0]$ becomes an \emph{invertible} simple
current that survives the strictly linear flow.

\paragraph{Factorisation of $W_0$.}
For even $d$ the root $\omega_d^{m}=-1$ is real, so in addition to $X=Y$ (from
$n=0$) the potential has the real root $X=-Y$ (from $n=m$):
\be
 W_0=X^{d}-Y^{d}=(X-Y)(X+Y)\prod_{n=1}^{m-1}\bigl(X^2-(\omega_d^{n}+\omega_d^{-n})XY+Y^2\bigr)\ ,
\ee
the product now running over the $m-1$ conjugate pairs $\{n,d-n\}$, $n=1,\dots,m-1$.

\paragraph{\boldmath The lines and the relevant gcd.}
The matrix factorisation of $[2\ell,0,0]$ is still given by $E_0=\prod_{|n|\le\ell}(X-\omega_d^{n}Y)$,
which for $\ell<m$ keeps the same form as in the odd case,
\be\begin{split}
 E_{[\ell]}&=(X-Y)\prod_{n=1}^{\ell}\bigl(X^2-(\omega_d^{n}+\omega_d^{-n})XY+Y^2\bigr)\ ,\\
 J_{[\ell]}&=(X+Y)\prod_{n=\ell+1}^{m-1}\bigl(X^2-(\omega_d^{n}+\omega_d^{-n})XY+Y^2\bigr)\ ,
\end{split}\ee
the extra factor $(X+Y)$, i.e.\ the $n=m$ root, now sitting in $J_{[\ell]}$.
The argument of Section~\ref{sec:norank1} needs the greatest common divisor of
$W_0$ and $W_1=X^{d-2}-Y^{d-2}$. For even $d$ both linear factors are common,
since both $X=\pm Y$ solve $X^{d}-Y^{d}=0$ \emph{and} $X^{d-2}-Y^{d-2}=0$,
\be
 \gcd(W_0,W_1)=(X-Y)(X+Y)=X^2-Y^2\ .
\ee
Pulling this out, $W(\lambda)=(X^2-Y^2)\,\widehat W(\lambda)$, and the
irreducible-piece argument goes through verbatim: each of $(X-Y)$, $(X+Y)$, and
$\widehat W(\lambda)$ must sit entirely inside $E(\lambda)$ or $J(\lambda)$. The
deformable (``trivial'') rank-1 factorisations are therefore those for which $E$
or $J$ lies in $\{1,\ X-Y,\ X+Y,\ X^2-Y^2\}$.

\paragraph{First order and the obstruction.}
The formulae of Sections~\ref{sec:1st-order} and \ref{sec:order2} are unchanged. In particular, the formula $\delta_n = \frac{1}{d} (\omega_d^n - \omega_d^{-n})$ remains the same, and therefore (\ref{sumdelta}) still holds, as do (\ref{sumsigma}) and (\ref{sumsigma1}).
The only relevant change is the range of $\ell$: for even $d=2m$, $\ell$ runs over $1\le\ell\le m-1$ ($L=2\ell\le k=2m-2$), and the factor
$\sin\bigl((2\ell+2)\varphi\bigr)=\sin\bigl((\ell+1)\pi/m\bigr)$ now \emph{vanishes
at the upper endpoint}:
\be
 S_1-S_3\neq0\quad\text{for }0<\ell<m-1\ ,\qquad S_1-S_3=0\quad\text{at }\ell=m-1\ .
\ee
The obstruction thus arises at second order for $0<\ell<m-1$, exactly as in the
odd case, while the top line $\ell=m-1$ is \emph{not} obstructed.

\paragraph{Interpretation of $\ell=m-1$.}
The unobstructed line is $[k,0,0]=[2(m-1),0,0]$, with quantum dimension
\be
 \gamma_{[k,0,0]}(0,0,0)=\frac{\sin\bigl(\pi(k+1)/(k+2)\bigr)}{\sin\bigl(\pi/(k+2)\bigr)}=1\ ,
\ee
i.e.\ it describes the invertible simple current generating the $\mathbb{Z}_2$ symmetry. Its
factorisation has $E_0=W_0/(X+Y)$ and $J_0=X+Y$. Since $(X+Y)$ divides both $W_0$
and $W_1$ (the latter because $d-2$ is even), it deforms \emph{exactly}, the series
terminating at first order,
\be
 E_{[m-1]}(\lambda)=E_0+\lambda\,\frac{W_1}{X+Y}\ ,\qquad J_{[m-1]}(\lambda)=X+Y\ ,\qquad
 E_{[m-1]}(\lambda)\,J_{[m-1]}(\lambda)=W_0+\lambda W_1\ ,
\ee
along the strictly linear flow. This is the even-$d$ counterpart of the trivial
cases $E,J\in\{1,X-Y\}$ of Section~\ref{sec:norank1}, and reflects that
$W(\lambda)=X^{d}+\lambda X^{d-2}$ is $\mathbb{Z}_2$-even, so the simple current is a
genuine symmetry of the deformed superpotential. The remaining lines
$0<\ell<m-1$ are non-invertible and obstructed exactly as before. The first
obstructed even case is $d=6$ ($k=4$, $\ell=1$), which is precisely the case in
which the fusion-ring argument of Section~\ref{sec:puzzle} is inconclusive, see
footnote~\ref{fn:k4}. For $d=4$ ($k=2$) the only
nontrivial integer-spin line is the simple current itself, so there is no obstruction, consistent with the endpoint vanishing
$S_1-S_3=0$ at $\ell=m-1$ above (for $d=4$, $m=2$, this is the single
line $\ell=1$).
\paragraph{Massive completion.}
The factorisation theorem of Appendix~\ref{app:factor} also carries over with only one change:
for even $d=2m$ the root $\omega_d^{m}=-1$ is real, so the grouping that leads to eq.~(\ref{B.3}), i.e.\ the grouping of
$\prod_{s=0}^{d-1}\bigl(X-(\omega_d^{s}\eta+\omega_d^{-s}a/\eta)\bigr)$ into conjugate
pairs $\{s,d-s\}$ now leaves \emph{two} singlets, $s=0$ and $s=m$, and only
$m-1$ pairs $j=1,\dots,m-1$. The $s=0$ singlet is $X-(\eta+a/\eta)=X-Y$ as before,
while the $s=m$ singlet is $X-(\omega_d^{m}\eta+\omega_d^{-m}a/\eta)=X+Y$. Both remain
undeformed. The pairs are unchanged, so the theorem reads, with $a=-\lambda/d$,
\be
 W(X,\lambda)-W(Y,\lambda)=(X-Y)(X+Y)\prod_{j=1}^{m-1}\Bigl(X^2-(\omega_d^{j}+\omega_d^{-j})XY+Y^2+a(\omega_d^{j}-\omega_d^{-j})^2\Bigr)\ .
\ee
The Chebyshev/Dickson potential $W(X,\lambda)=\xi^{d}+a^{d}/\xi^{d}$ is unchanged.  It
has $d-1$ massive vacua (the roots of $U_{d-1}$), and Gepner's identification
with the $\mathfrak{su}(2)_{d-2}$ chiral ring holds verbatim.

\paragraph{Appendix~\ref{app:result}.}
The proof in Appendix~\ref{app:result} applies also directly for even $d$. Again, the only change is that
the middle block degenerates. At $r=m$ one has
$(\omega_d^{m}-\omega_d^{-m})^2=0$ and $\omega_d^{m}+\omega_d^{-m}=-2$, so
$g_m=(X+Y)^2$, and consequently
\be
E_{[m]}=(X+Y) \bigl( W(X,\lambda)-W(Y,\lambda) \bigr) \ ,
\ee
rather than $E_{[m]}=W(X,\lambda)-W(Y,\lambda)$, as in the odd case. This does not enter the
level-truncation argument, which only uses the reflection \eqref{reflection}, valid for either parity,
together with the fact that for even $d$ the map $\ell\mapsto d-1-\ell$ has no fixed point. 
Hence eq.~\eqref{resultant} holds unchanged.

\bibliography{reference}
\bibliographystyle{utphys}

\end{document}